\documentclass[fleqn, usenatbib]{mnras}

\usepackage{adjustbox}
\usepackage[T1]{fontenc}
\usepackage{ae,aecompl}
\usepackage{longtable}
\usepackage{graphicx}   % Including figure files
\usepackage{amsmath}    % Advanced maths commands
\usepackage{amssymb}    % Extra maths symbols

\title[Optical GeV connection in BL Lacs objects]{\textbf{Correlation between optical 
and $\gamma$ -ray flux variations in BL Lacs}}

\author[Bhoomika et al.]{Bhoomika Rajput$^{1}$\thanks{E-mail: bhoomika@iiap.res.in},
Zahir Shah$^{2}$,
C. S. Stalin$^{1}$,
S. Sahayanathan$^{3}$,
Suvendu Rakshit$^{4,5}$
\\
$^{1}$Indian Institute of Astrophysics, Block II, Koramangala, Bangalore 560034, India\\
$^{2}$Inter-University Center for Astronomy and Astrophysics, Post Bag 4, Ganeshkhind, Pune, India - 411007\\
$^{3}$Astrophysical Sciences Division, Bhabha Atomic Research Centre, Mumbai, India\\
$^{4}$Finnish Centre for Astronomy with ESO (FINCA), University of Turku, Quantum, Vesilinnantie 5, 20014 University of Turku, Finland\\
$^{5}$ Aryabhatta Research Institute for Observational Sciences (ARIES), Nainital, India}
\date{Last updated 2020 Dec 24; in original form 2020 December 24}

\begin{document}
\label{firstpage}
\pagerange{\pageref{firstpage}--\pageref{lastpage}}
\maketitle

\begin{abstract}
We report here results on the analysis of correlated flux variations between the optical and GeV $\gamma$-ray bands in three bright BL Lac objects, namely AO\, 0235+164, OJ 287 and PKS 2155$-$304. This was based on the analysis of about 10 years of data from the {\it Fermi} Gamma-ray Space Telescope covering the period between 08 August 2008 to 08 August 2018 along with optical data covering the same period. For all the sources, during the flares analysed in this work, the optical and $\gamma$-ray flux variations are found to be closely correlated. From broad band spectral energy distribution modelling of different epochs in these sources using the one zone leptonic emission model, we found that the optical-UV emission is dominated by synchrotron emission from the jet. The $\gamma$-ray emission in the low synchrotron peaked sources AO\, 0235+164 and OJ 287 are found to be well fit with external Compton (EC) component, while, the $\gamma$-ray emission in the high synchrotron peaked source PKS 2155$-$304 is well fit with synchrotron self Compton component. Further we note that the $\gamma$-ray emission during the high flux state of AO 0235+164 (epochs A and B) requires seed photons from both the dusty torus and broad line region, while the $\gamma$-ray emission in OJ 287 and during epochs C and D of AO\,0235+164 can be modelled by EC scattering of infra-red photons from the torus.

%We report here results on the analysis of correlated flux variations between the optical and GeV $\gamma$-ray bands in three bright BL Lac objects, namely AO 0235+164, OJ 287 and PKS 2155$-$304. This was based on the analysis of about 10 years of data from the {\it Fermi} Gamma-ray Space Telescope covering the period between 08 August 2008 to 08 August 2018 along with optical data covering the same period. For all the sources, during the flaresanalysed in this work, the optical and $\gamma$-ray flux variations are found to be closely correlated. From broad band spectral energy distribution modelling of different epochs in those sources using the one zone leptonic emission model, we found that the optical-UV emission is dominated by synchrotron emission from the jet. The $\gamma$-ray emission in the low synchrotron peaked sources AO 0235+164 and OJ 287 are found to be well fit with external Compton component, while, the $\gamma$-ray emission  in the high synchrotron peaked source PKS 2155$-$304 is well fit with synchrotron self Compton component. In the sources OJ 287 and PKS 2155$-$304 we found hard $\gamma$-ray spectrum, not explainable by the one zone leptonic emission model; whereas, in the case of AO 0235+164 the contributions from of SSC and EC mechanisms are capable to explain the hard $\gamma$-ray spectrum. The flux variability behaviours of the three sources during different epochs could be explained by changes in the jet parameters of the sources such as the bulk Lorentz factor, magnetic field and the break energy.
\end{abstract}



%%%%%%%%%%%%%%%%%%%%%%%%%%%%%%%%%%%%%%%%%%%%%%%%%%

%%%%%%%%%%%%%%%%% BODY OF PAPER %%%%%%%%%%%%%%%%%%

% The MNRAS class isn't designed to include a table of contents, but for this document one is useful.
% I therefore have to do some kludging to make it work without masses of blank space.
%\begingroup
%\let\clearpage\relax
%\tableofcontents
%\endgroup
%\newpage

\section{Introduction}
Blazars, a class of active galactic nuclei (AGN) and one of the most luminous 
objects (10$^{42}$ $-$ 10$^{48}$ erg s$^{-1}$) in the extragalactic $\gamma$-ray sky, are 
believed to be powered by accretion of matter onto super massive black holes 
with masses between $\sim$ 10$^8$ $-$ 10$^{10}$ M$_{\odot}$ situated 
at the centers of galaxies \citep{1969Natur.223..690L,1973A&A....24..337S}. 
They constitute about 2/3 of the point sources in the eight year {\it Fermi} 
catalog \citep{2020ApJS..247...33A}.
These objects have relativistic jets pointed close (within a few degrees) 
to the observer. Non-thermal emission dominates the radiation from these sources 
\citep{1993ARA&A..31..473A,1995PASP..107..803U} and spans the entire electromagnetic 
spectrum from low energy radio to high energy $\gamma$-rays. One of the defining 
characteristics of blazars is that they show rapid flux variations over the 
complete accessible electromagnetic spectrum on a wide range of time scales from
minutes to years. In addition to flux variability, they are strongly 
polarized and also show 
polarization variations \citep{1966ApJ...146..964K,1980ARA&A..18..321A,2005A&A...442...97A,2017ApJ...835..275R}. They have a compact radio 
structure often with core jet morphology, 
flat radio spectra and also show superluminal 
motion \citep{1995ARA&A..33..163W,1997ARA&A..35..445U,2014A&ARv..22...73F,2016ARA&A..54..725M}. Traditionally blazars are divided into two 
classes namely flat spectrum radio quasars (FSRQs) and BL Lac objects (BL Lacs). This 
classification is based on the presence or absence of emission lines in their
optical/infrared (IR) spectra with FSRQs having strong emission lines and BL Lacs
either having no or weak emission lines with equivalent widths 
$<$ 5 \AA~  \citep{1995PASP..107..803U}. A physical distinction between FSRQs and BL Lacs 
has been proposed by \cite{2011MNRAS.414.2674G} according to which the ratio of 
the luminosity of the broad line region ($L_{BLR}$) to the Eddington 
luminosity ($L_{Edd}$) is $>$ 5 $\times$ 10$^{-5}$ in the case of FSRQs.

\begin{table*}
\caption{Details of the objects studied in this work. The average $\gamma$-ray fluxes during the period 2008-2018 August between 100 MeV to 300 GeV are in units of 10$^{-7}$ ph cm$^{-2}$ s$^{-1}$ and $\Gamma$ is the photon index in the $\gamma$-ray band.}
\begin{tabular} {lcccccr} \hline
Name           & 3FGL Name       & $\alpha_{2000}$ & $\delta_{2000}$  & $z$ & $\Gamma$ & $\gamma$-ray flux \\ \hline
AO 0235+164    &  3FGL J0238.6+1636  & 02:38:38.9 &   +16:36:59 &  0.940 & 2.06 & 1.41 \\
OJ 287         &  3FGL J0854.8+2006  & 08:54:48.9 &   +20:06:31 &  0.306 & 2.12 & 0.99 \\
PKS 2155$-$304 &  3FGL J2158.8-3013  & 21:58:52.0 & $-$30:13:32 &  0.116 & 1.75 & 1.12 \\ \hline
\end{tabular}
\label{table-1}
\end{table*}

The spectral energy distribution (SED) of blazars in the $log\nu F_{\nu}$ v/s $log\nu$ 
representation has two dominant humps, the low energy hump peaks in the UV/optical 
wavelengths and the high energy hump peaks in the $\gamma$-ray 
\citep{1998MNRAS.299..433F,2016ApJS..224...26M} wavelengths. The 
low energy component in the broad band SED of blazars is well understood and it is attributed to synchrotron emission due to relativistic electrons moving in the magnetic field of the jet. However, the nature of the high energy emission from the jets of blazars is still debated. A likely process
is inverse Compton (IC) scattering of low energy photons \citep{2010ApJ...716...30A}, these seed photons can be 
photons from the jet (synchrotron self Compton; SSC; \citealt{1981ApJ...243..700K,1985ApJ...298..114M,1989ApJ...340..181G}) or photons external to the jet
such as the broad line region (BLR) \citep{1996MNRAS.280...67G}, torus \citep{2000ApJ...545..107B,2008MNRAS.387.1669G} and the accretion
disk \citep{1997A&A...324..395B}. The other possible mechanisms for the high 
energy emission in blazars are hadronic processes \citep{2013ApJ...768...54B}, 
that include proton synchrotron \citep{2000NewA....5..377A} or
photon pion processes \citep{1993A&A...269...67M}. Based on the position of the peak
frequency ($\nu_p$) of the synchrotron emission in their 
broad band SED \citep{1995ApJ...444..567P,2010ApJ...716...30A}, blazars 
are subdivided into low synchrotron peaked blazars (LSP, $\nu_p  < 10^{14}$ Hz), intermediate synchrotron peaked blazars (ISP, $10^{14} Hz < \nu_p < 10^{15} Hz$) and high synchrotron
peaked blazars (HSP, $\nu_p > 10^{15} Hz$). One of the approaches to understand 
the physical processes that
contribute to the observed emission over a wide range of wavelengths and constrain
the leptonic versus hadronic emission process in blazars is by detailed
broad band SED modelling. Nevertheless such studies are highly sensitive to the adopted models
and the capability to acquire simultaneous observations covering a broad range
of wavelengths. Also, often the observed SED of blazars can be modelled satisfactorily by
leptonic \citep{2015ApJ...803...15P}, hadronic \citep{2013ApJ...768...54B} and 
lepto-hadronic models \citep{2016ApJ...817...61P}.

An alternative to this SED modelling approach is via correlated studies of flux
variations between low energy (optical) and high energy ($\gamma$-ray) bands. In the
leptonic scenario of emission from blazar jets, close correlation between the 
flux variations in the low energy optical and the high energy $\gamma$-ray bands
is expected \citep{2007Ap&SS.309...95B}. In the hadronic model of emission from blazar jets, 
as the optical emission is dominated by electron synchrotron from jets and the 
$\gamma$-ray emission from proton synchrotron, a correlation between
optical and GeV $\gamma$-rays need not be expected \citep{2001APh....15..121M}. 
Therefore, by a systematic investigation of the correlation between the optical and 
GeV $\gamma$-ray flux variations, one can put constraints on the leptonic v/s hadronic
emission from blazar jets. Studies available in the literature have found  (a) close
correlation between optical and $\gamma$-ray flares \citep{2009ApJ...697L..81B}, (b) optical
flares without the corresponding $\gamma$-ray 
flares \citep{2013ApJ...763L..11C,2014ApJ...797..137C} and (c) $\gamma$-ray flares without
optical counterparts \citep{2013ApJ...779..174D,2015ApJ...804..111M}. To look for the 
prevalence of anomalous variations between the optical and GeV $\gamma$-rays we 
have carried out a systematic investigation of the $\gamma$-ray and optical flux variations in blazars.
Results on the FSRQ category of blazars were reported by 
\cite{2019MNRAS.486.1781R,2020MNRAS.tmp.2558R}.
We present here the results of the BL Lac category of blazars. Details on the selection
of the objects are given in Section 2. In Section 3 we give the details of the data used
in this work, the analysis is discussed in Section 4 followed by the results in Section 5.
We summarize our findings in the final Section.

\begin{table*}
\caption{Summary of the epochs considered for detailed light curve analysis, 
SED modelling and spectral analysis. The $\gamma$-ray fluxes are between 100 
MeV to 300 GeV and in units of 10$^{-6}$ ph cm$^{-2}$ s$^{-1}$ and the optical 
fluxes are in units of 10$^{-11}$ erg cm$^{-2}$ s$^{-1}$. The entry OG 
in the final column indicates that the optical and $\gamma$-ray flares are 
	correlated while Q refers to the quiescent level. The mean 
	$\gamma$-ray flux during the quiescent state of AO 0235+164, OJ 287 and PKS 2155$-$304 
	are 0.11 $\pm$ 0.11, 0.04 $\pm$ 0.06 and 0.08 $\pm$ 0.06  $\times$ 10$^{-6}$ ph cm$^{-2}$ s$^{-1}$ respectively. 
	Similarly in the optical band, the mean brightness during the quiesscent states
	of AO 0235+164, OJ 287 and PKS 2155$-$304 are 0.07 $\pm$ 0.02, 1.62 $\pm$ 0.46 and
	5.82 $\pm$ 0.61 $\times$ 10$^{-11}$ erg cm$^{-2}$ s$^{-1}$ respectivey.}. 
\begin{tabular} {lccccccccl} \hline
              & \multicolumn{2}{c}{MJD}  & \multicolumn{2}{c}{Calender date}   & \multicolumn{2}{c}{Peak Flux} & \multicolumn{2}{c}{Peak/Mean Flux}  &  \\
Name/ID       & Start    & End           & Start          & End                &  $\gamma$      & Optical  &  $\gamma$      & Optical &  Remark  \\ \hline    
AO 0235+164      &          &               &                &                 &                 &                 &                 &                 &   \\
A                &  54720   & 54740  & 11-09-2008 & 01-10-2008 & 0.86$\pm$0.14 & 2.26$\pm$0.01 & 7.82$\pm$1.39 & 32.29$\pm$0.72 & OG \\
B                &  54743   & 54763  & 04-10-2008 & 24-10-2008 & 1.41$\pm$0.20 & 1.57$\pm$0.01 & 12.82$\pm$2.93  & 22.43$\pm$0.50 & OG \\
C                &  55100   & 55200  & 26-09-2009 & 04-01-2010 & ---                 & ---                 &  ---                  & ---                  & Q  \\
D                &  57040   & 57060  & 18-01-2015 & 07-02-2015 & 0.22$\pm$0.10 & 0.67$\pm$0.01 & 2.00$\pm$0.30 & 9.57$\pm$0.21  & OG \\ \hline
OJ 287           &          &        &            &            &                     &                 &                  &                  &    \\
A                &  54870   & 54970  & 08-02-2009 & 19-05-2009 & --- & --- & --- & ---     & Q   \\
B                &  55127   & 55147  & 23-10-2009 & 12-11-2009 & 0.50$\pm$0.18 & 5.88$\pm$0.02 & 12.50$\pm$2.37  & 3.63$\pm$1.67 & OG  \\ 
C                &  56735   & 56755  & 19-03-2014 & 08-04-2014 & 0.59$\pm$0.18 & 2.69$\pm$0.01 & 14.75$\pm$2.80  & 1.66$\pm$0.76 & OG  \\ 
D                &  57350   & 57370  & 24-11-2015 & 14-12-2015 & 0.47$\pm$0.14 & 8.20$\pm$0.02 & 11.75$\pm$1.79  & 5.06$\pm$2.33 & OG \\  \hline
PKS 2155$-$304   &          &        &            &                                  &                 &                 &                  &                 &    \\ 
A                &  55740   & 55840  & 28-06-2011 & 06-10-2011      & --- & ---&  --- & ---     & Q \\ 
B                &  56790   & 56810  & 13-05-2014 & 02-06-2014      & 1.04$\pm$0.20 & 10.70$\pm$0.02 & 13.00$\pm$2.72 & 1.84$\pm$1.12 & OG \\ \hline
\end{tabular}
\label{table-2}
\end{table*}

\section {Sample}
The selection of BL Lac objects for the analysis of correlated variations between the 
optical and GeV bands was done in the same manner as it was carried out for 
FSRQs \citep{2020MNRAS.tmp.2558R}
and we describe them in brief below. Firstly, we 
selected all the sources classified as BL Lacs in the third catalog of AGN detected by 
the large area telescope (LAT) onboard the {\it Fermi} Gamma Ray Space
Telescope, hereinafter {\it Fermi} (3LAC; \citealt{2015ApJ...810...14A}). 
For the selected BL Lacs we looked into their one day binned $\gamma$-ray light 
curves for the period of 10 years and then we selected those sources that have 
atleast one flare in the $\gamma$-ray band with the $\gamma$-ray flux exceeding 10$^{-6}$ 
photons cm$^{-2}$ s$^{-1}$.  With this criteria we arrived at a sample
of 21 BL Lacs. As the motivation of this work is to look for any correlation
between optical and GeV flux variations, for those 21 BL Lacs, 
we looked in the archives of the Small and Moderate Aperture Research 
Telescope System\footnote{http://www.astro.yale.edu/smarts/glast/home.php} (SMARTS, \citealt{2009ApJ...697L..81B}) for the availability of 
optical and infrared (IR) data overlapping the $\gamma$-ray data. Of the 21 sources, for 6 
objects we found optical and IR data overlapping with the $\gamma$-ray light curves.
Thus our sample of BL Lac objects  with overlapping optical, IR  and 
$\gamma$-ray light curves consists of six sources, namely AO 0235+164, 
PKS 0301$-$243, PKS 0426$-$380, PKS 0537$-$441, OJ 287 and PKS 2155$-$304. Ordering in terms of the total epochs of optical and IR observations available on these objects in the SMARTS archives, in this work, we are presenting the results of the top three sources namely AO 0235+164, OJ 287 and PKS 2155$-$304. The details of the three 
sources are given in Table \ref{table-1} and a brief description of them 
is given below. However, the reviewed references are only a small sub-set 
of the large body of literature available on the multi-wavelength studies and 
modelling on these sources.

\subsection{AO 0235+164} 
AO 0235+164 was first classified as a BL Lac object 
based on its variability and featureless optical spectrum \citep{1975ApJ...201..275S}. 
\cite{1987ApJ...318..577C} measured the redshift of the object at $z$ = 0.94. 
It has shown violent variations across the electromagnetic 
spectrum that includes optical, X-rays and $\gamma$-rays 
\citep{2001A&A...377..396R,2009A&A...507..769R,2010ApJ...716...30A}. 
The high optical flux variability shown by the source during December 2006 is also accompanied by high optical polarization (30\% - 50\%) 
\citep{2008ApJ...672...40H}. 
\cite{2001A&A...377..396R} found the quasi-periodic behaviour in radio outburst 
with a  periodicity of $\sim$ 5.7 years. It is classified as a LSP blazar 
\citep{2015ApJ...810...14A} and its $\gamma$-ray spectrum is well fit by a 
log-parabola function \citep{2015ApJS..218...23A}. From detailed multi-wavelength 
observations of the source spanning about six months and including observations
from the radio to the $\gamma$-ray bands, \cite{2012ApJ...751..159A} 
found the $\gamma$-ray activity to be well correlated with the optical/IR flares. 
They also found the broad band SED to be well explained by leptonic models with 
the seed photons for the IC scattering from the torus.

\subsection{OJ 287} OJ 287 first identified as a blazar in 1967 
by \cite{1967AJ.....72..757D} is a LSP blazar \citep{2015ApJ...810...14A} at 
redshift $z$ = 0.306. In the long term optical light curve, a periodicity of $\sim$ 12 
years was observed which in the binary super massive black hole model is attributed 
to the secondary super massive black hole striking the accretion disk 
around the primary super massive black hole 
\citep{1988ApJ...325..628S}. In addition to flux variations in the optical band 
\citep{2017ApJ...835..275R,2017ApJ...844...32P}, it has also shown polarization 
variations \citep{2017ApJ...835..275R}. Flux variations are also seen in the GeV
$\gamma$-ray energy band \citep{2011ApJ...726L..13A}. The broadband spectral analysis 
of the source at various activity levels points to the high energy hump explained by 
inverse Compton scattering of both emission line photons \citep{2018MNRAS.473.1145K} 
as well as the torus \citep{2013MNRAS.433.2380K}.

\subsection{PKS 2155$-$304} PKS 2155$-$304 was first discovered in the Parkes 
survey \citep{1974AuJPA..32....1S} and identified as a BL Lac object by 
 \cite{1980ApJS...43...57H}. It is a HSP blazar at a redshift of $z$ =0.116 
 \citep{1984ApJ...278L.103B}, and its $\gamma$-ray spectrum is consistent with a 
log parabola function \citep{2015ApJS..218...23A}. Correlated optical and $\gamma$-ray 
flare was seen in 2014, while in 2016, there was a large optical flare with no
corresponding flare both in the GeV band as well as at very high energies
\citep{2019arXiv191201880W}. Simultaneous observations of the source in 2008, 
showed evidence of correlated flux variations between the optical and the very high energy (VHE) $\gamma$-ray bands, however, the 
increased flux in the optical band has no correspondence with the X-ray and the  GeV flux 
consistent with being constant \citep{2009ApJ...696L.150A}.
Quasi-periodic variations in the optical emission with a time scale of 
317$\pm$12 days \citep{2014RAA....14..933Z} and in the GeV emission with a time 
scale of 1.74$\pm$0.013 years are known in PKS 2155$-$304 \citep{2017ApJ...835..260Z}.

\begin{figure*}
\hspace*{-3cm}
\includegraphics[width=1.3\textwidth]{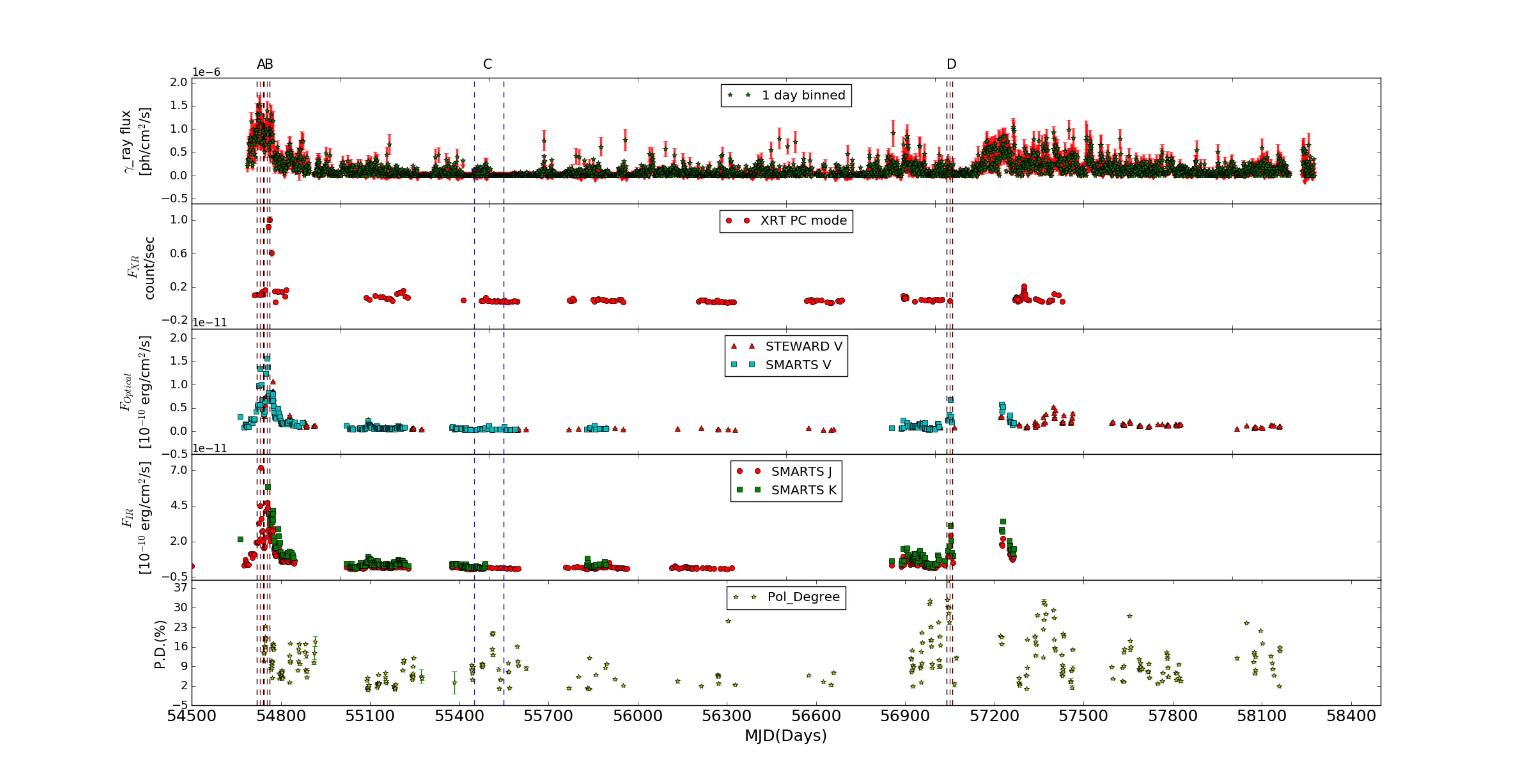}
\caption{Multi-wavelength light curves of the source AO 0235+164. From the top, panels refer to the one day binned $\gamma$-ray light curve, the X-ray light curve, the optical light curve, the IR light curves and the degree of optical polarization. The vertical red lines refer to the optical flare peaks and $\gamma$-ray flare, and the two vertical black lines indicate a period of 10 days each on either side of the peak of the flare. The two vertical blue lines are for the period of 100 days and shown correspond to the quiescent period. The upper limit points, which are defined for TS $<$ 9 are shown with vertical arrow in the one day binned $\gamma$-ray light curve.}
\label{figure-1}
\end{figure*}  

\begin{figure*}
\begin{center}$
\begin{array}{rr}
\includegraphics[width=75mm,height=85mm]{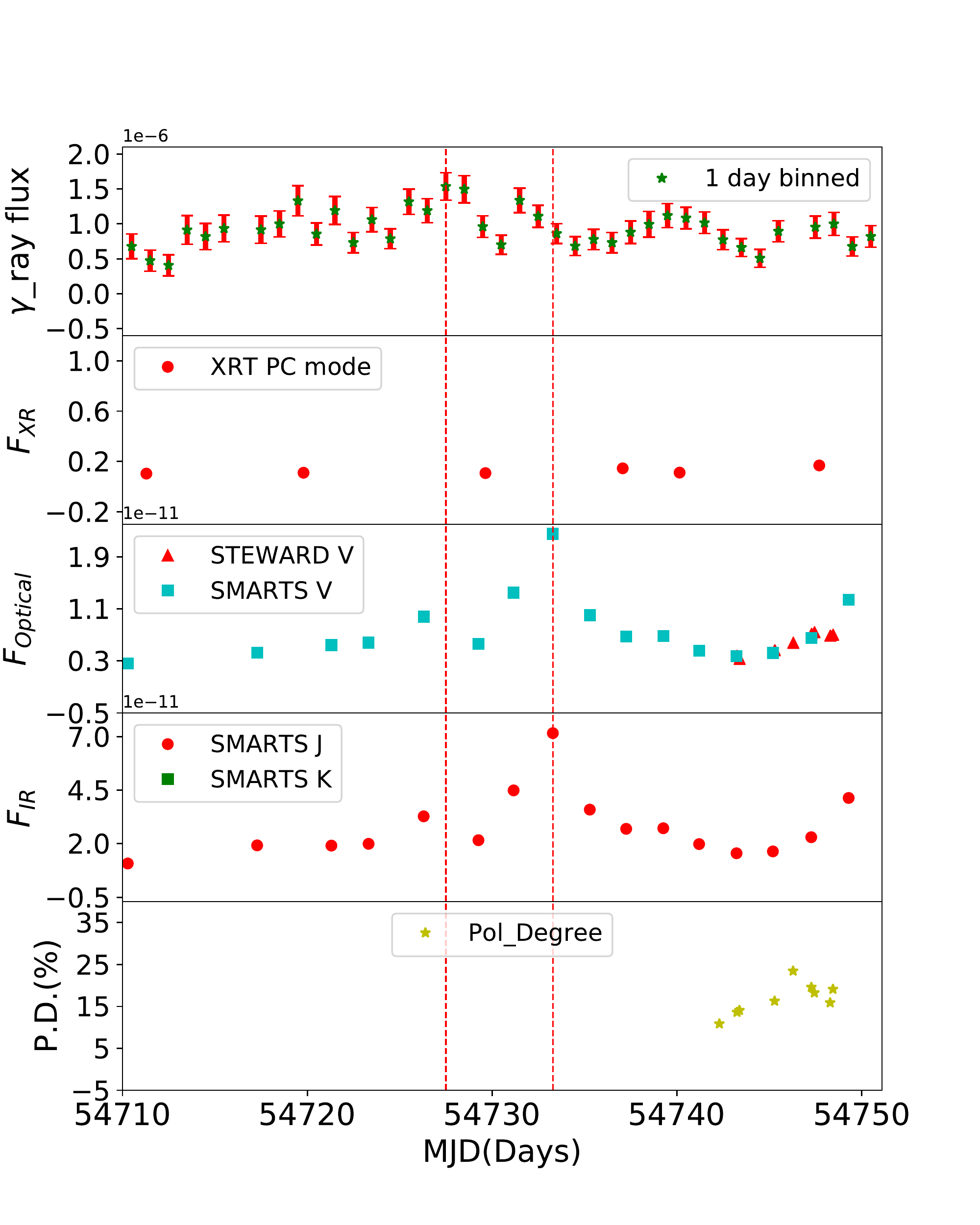}
\includegraphics[width=75mm,height=85mm]{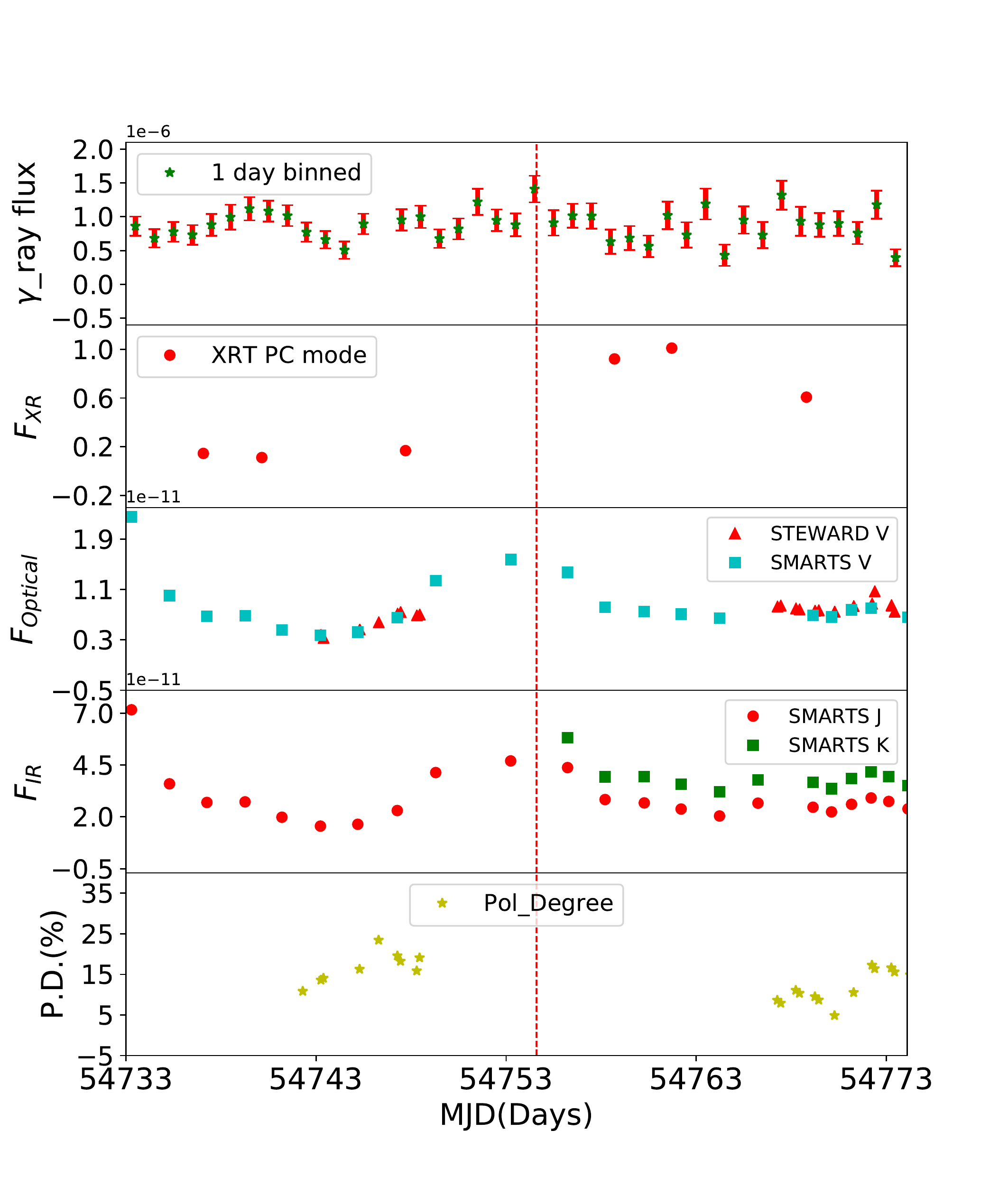}
\end{array}$
\end{center}
\begin{center}$
\begin{array}{rr}
\includegraphics[width=75mm,height=85mm]{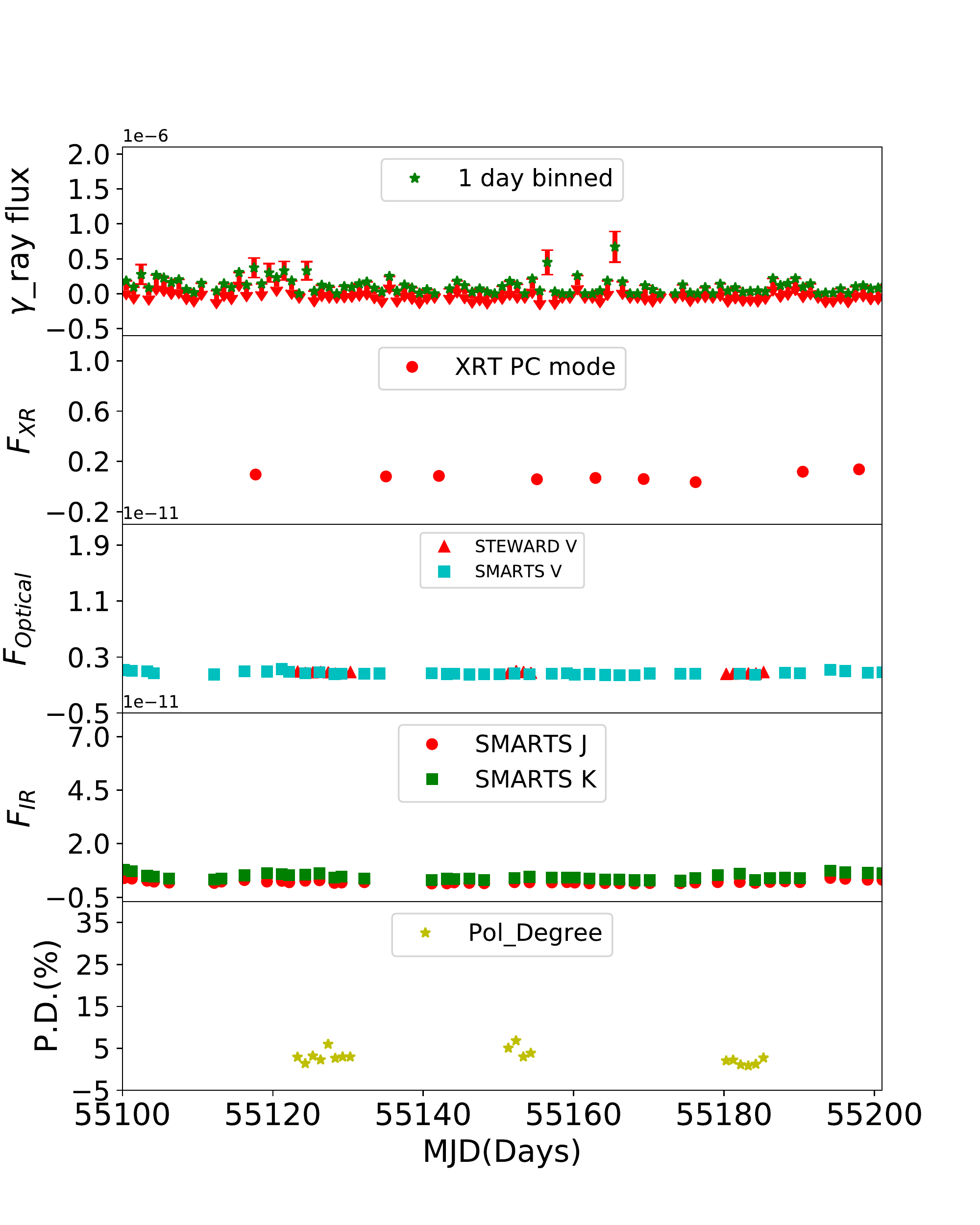}
\includegraphics[width=75mm,height=85mm]{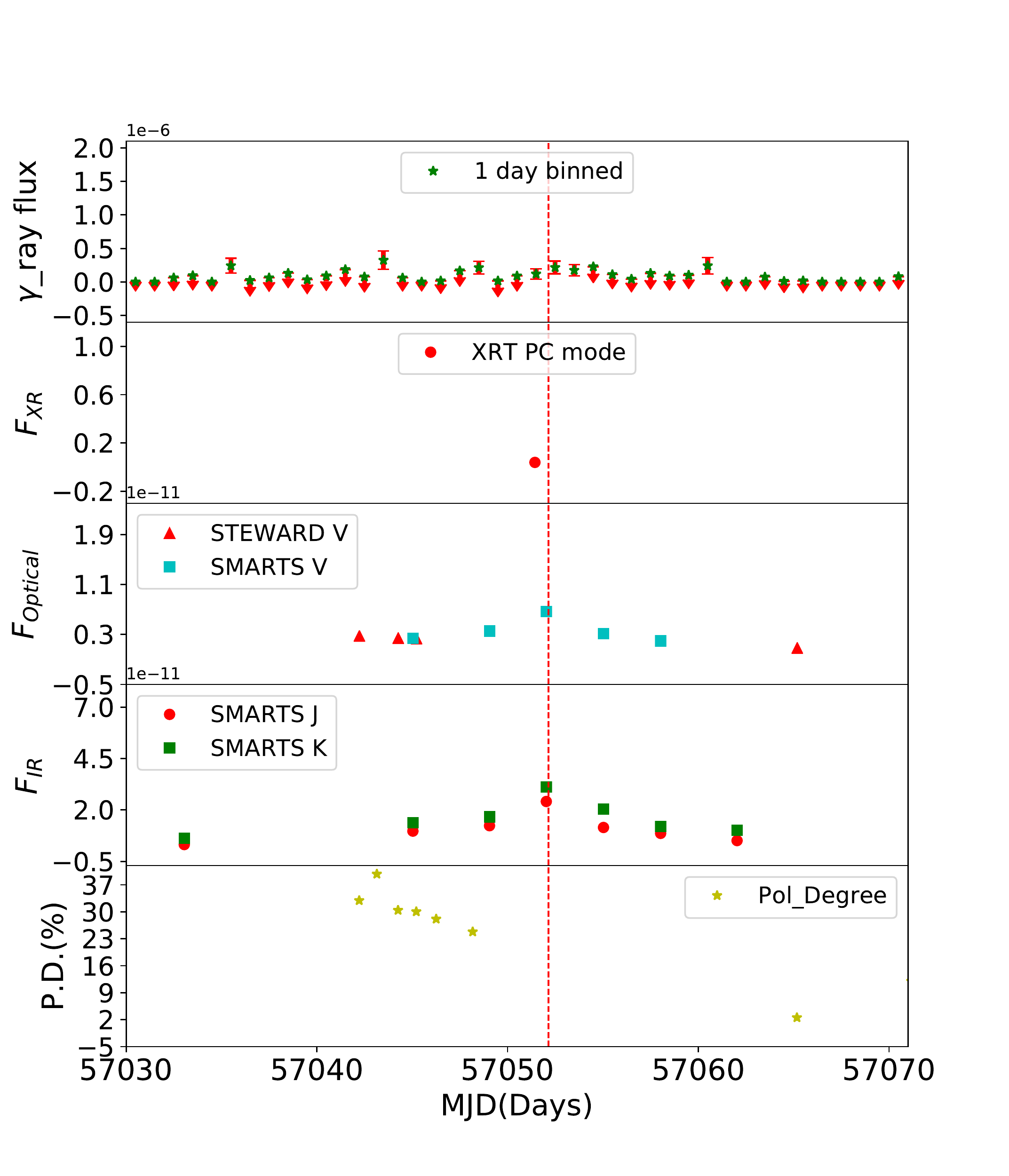}
\end{array}$
\end{center}
\caption{Multi-wavelength light curves for the selected epochs of the source AO 0235+164. 
Epoch A is in the top left panel and epoch B is shown in the top right panel. The bottom
left and right panels show the light curves for epochs C and D respectively. The dashed lines show the peak of the optical and GeV flare.}
\label{figure-2}
\end{figure*}

\section{Reduction of Multi-wavelength data} \label{sec:data}
The main motivation in this work is to look for correlated variations between the 
optical and GeV $\gamma$-ray bands. This demands the availability of data in both
optical and $\gamma$-ray bands. Furthermore, for broad band SED modelling of few selected
epochs, data from other bands such as the IR, ultraviolet (UV) and X-ray 
are also needed. Therefore, we used all publicly available data in the IR, optical, UV, 
X-rays and $\gamma$-rays that span the period between 08 August 2008 and 08 August 2018.
Optical polarimetric data if available during the above period were also used.

\subsection{$\gamma$-ray data}
The $\gamma$-ray data covering the period 08 August 2008 - 08 August 2018 were
from the LAT instrument onboard the {\it Fermi} Gamma Ray 
Space Telescope. We generated one day binned $\gamma$-ray light curve 
using {\it Fermipy} \citep{2017arXiv170709551W}. We used Pass 8 data for the analysis 
where the photon-like events are classified as 'evclass=128, evtype=3' with energy 
range 0.1$\leqslant$E$\leqslant$300 GeV. A circular region of interest (ROI)  
of $15^{\circ}$ was chosen with zenith angle cut $90^{\circ}$ in order to remove earth 
limb contamination. We used the isotropic model "iso$\_$P8R2$\_$SOURCE$\_$V6$\_$v06" 
and the Galactic diffuse emission model "gll$\_$iem$\_$v06" for the analysis. 
The recommended criteria "(DATA$\_$QUAL>0)\&\&(LAT$\_$CONFIG==1)" was used for 
the good time interval selection. In the one day binned light curve the source
is considered to be detected if the test statistics (TS) $>$ 9. This corresponds 
to a 3$\sigma$ detection \citep{1996ApJ...461..396M}. Epochs with 
TS $<$ 9 are shown as upper limits in the one day binned $\gamma$-ray light 
curves. However, as points with TS $<$ 9 are also detections, but with 
significance $<$ 3$\sigma$, they were used in the  calculation of the average 
brightness of the sources in their faint states.

\subsection{X-ray data}
For X-rays we used data from the {\it Swift}-XRT telescope that covers the energy range 
of 0.3 $-$ 10 keV \citep{2005SSRv..120..165B,2004ApJ...611.1005G} for the period 2008 August
to 2018 August. This was taken from the archives of 
HEASARC\footnote{https://heasarc.gsfc.nasa.gov/docs/archive.html}. 
We analyzed the data with the default parameter settings as suggested by 
the instrument pipeline. For generation of the light curves, we used data from
both the window timing (WT) and photon counting (PC) mode, however, for 
spectral analysis we used PC mode data for the sources AO 0235+164 and 
OJ 287 for both the quiescent and flaring states, while for the source 
PKS 2155$-$304, we used
WT mode data for the quiescent state and PC mode data for the flaring state. We 
processed the data  with the {\tt xrtpipeline} task using the latest 
CALDB files available with version HEASOFT-6.24. We used the standard grade 
selection 0-12. We extracted the source spectra from a circular region of 
radii 60$^{\prime\prime}$, whereas, the background spectra were selected from the region of 
radii 80$^{\prime\prime}$ away from the source for PC mode. For WT mode, for the source we used a 
circular region of 60$^{\prime\prime}$ radii and for the background we used the 
region between circular radii of 80$^{\prime\prime}$ and 120$^{\prime\prime}$ centered around the source. We 
combined the exposure map using the tool {\tt XIMAGE} and 
to create the ancillary response files we used {\tt xrtmkarf}. We used an absorbed 
simple power law model with the Galactic neutral hydrogen column density of 
$N_{H}$=6.59$\times$ $10^{20}$ cm$^{-2}$, 2.38$\times$ $10^{20}$ cm$^{-2}$ and 
1.29$\times$ $10^{20}$ $cm^{-2}$ \citep{2005A&A...440..775K} for the sources 
AO 0235+164, OJ 287 and PKS 2155$-$304 respectively to perform the fitting within 
XSPEC \citep{1996ASPC..101...17A}. Within XSPEC, we adopted $\chi^2$ statistics and the
calculated uncertainties are at the 90\% confidence level.

\subsection{UV-optical and IR data}
For data in the UV and optical bands we used observations from {\it Swift}-UVOT.
The data from {\it Swift}-UVOT were analysed using the online 
tool\footnote{https://www.ssdc.asi.it/cgi-bin/swiftuvarchint}. In addition to the optical data from {\it Swift}-UVOT,
we also used optical data in the V-band from both SMARTS and the Steward Observatory, 
while infrared observations in the J and K-bands were taken from SMARTS \citep{2012ApJ...756...13B}. Corrections due to galactic
absorption were applied to the UV, optical and IR points for SED analysis. These magnitudes were not
corrected for contribution from host galaxy. The galactic absorption corrected magnitudes
were converted to fluxes using the zero points in \cite{1979PASP...91..589B} 
and \cite{2011AIPC.1358..373B}. Optical 
polarization data  was taken from the Steward Observatory \citep{2009arXiv0912.3621S} \footnote{http://james.as.arizona.edu/$\sim$psmith/Fermi}.

\section{Analysis}
\subsection{Multi-wavelength light curves}
The motivation behind this work is to examine the presence and/or absence
of correlated flux variations between optical and $\gamma$-ray bands. This requires
firstly identification of epochs where the optical and $\gamma$-ray flux variations
are correlated or uncorrelated. An automated procedure to identify this 
was hindered due to gaps and less number of data points in the optical light 
curves. Therefore, flares were first identified visually. Once those flares 
were identified expanded light curves were generated for a duration of 20 days 
centered on the optical or $\gamma$-ray flare. This is to make sure the 
availability of data at multiple wavelengths that are needed for broad band 
SED modelling. This condition of data availability at multiple wavelengths 
lead to the identification of few epochs for each object. For those epochs we 
carried out a quantitative assessment for the presence and/or absence of 
correlation between optical and $\gamma$-ray waveband through statistical 
analysis. For this we calculated over the 20 day period (a) the mean of the 
optical and $\gamma$-ray fluxes (that includes detections with TS $<$ 9)
during the quiescent periods, (b) peak 
of the optical and $\gamma$-ray flares and (c) the ratio of the peak of the 
optical/$\gamma$-ray flares to their mean flux levels at the
quiescent epochs. These are given in 
Table \ref{table-2}. Using the ratio we concluded
on the presence or absence of optical - $\gamma$-ray correlation. This was also cross checked by carrying out correlation 
function anaysis between the optical and $\gamma$-ray light curves
using the discrete correlation function  (DCF) method of 
\cite{1988ApJ...333..646E}.  Though the correlation functions are noisy due 
to the sparseness of the data at each epoch, the DCF analysis indicates
that the optical and $\gamma$-ray light curves of all the identified
epochs are correlated with lags consistent with zero. The details 
of the epochs selected for each object based on the above are further 
described below.
\subsubsection{\bf AO 0235+164}
Multi-wavelength light curves from {\it Fermi}, {\it Swift}-XRT, SMARTS and
Steward observatory are shown in Fig \ref{figure-1}. Two major flares in the optical/IR bands 
are evident in the light curves. X-ray and $\gamma$-rays too have measurements 
simultaneous to the flares in the optical/IR. Close inspection of the light 
curves indicates that these large optical flares are composed of many short 
term flux variations. We identified four epochs in this source namely A,B,C 
and D for studying the correlations between optical and GeV flux variations. 
Table \ref{table-2} provides a summary of these epochs 
and the multi-wavelength light curves covering a 40 day period for each epoch (except for the quiescent epoch which is 100 days) are given in Fig. \ref{figure-2}.  However, for SED analysis during flaring 
epochs, data covering a period of 20 days centered around the flare was used.
The details of those four epochs are given below.

\noindent {\bf Epoch A:} During this epoch, the $\gamma$-ray and optical flares 
have increased by a factor of $\sim$32 and $\sim$8 respectively relative to 
their quiescent epochs, however, the amplitude of IR variation is larger
than the variation in the optical. During this epoch, there are two 
optical flares, a major one around MJD 54733 and a minor one around MJD 54726.
For the peak of the $\gamma$-ray flux during this epoch, we considered
the one that coincides with the major optical flare.
The nature of X-ray flux variations during the peak of the optical flare 
could not be ascertained due to the lack of X-ray flux measurements. No optical polarization measurements were available during the 20 days period of this 
epoch. DCF analysis between optical and $\gamma$-ray light curves
during this epoch shows that both are correlated with a lag of 4.84$^{+1.8}_{-6.24}$ days. However, due to the sparseness of the data in the optical band, uncertainty is high. We conclude that in this epoch the optical and GeV flares are correlated.

\noindent {\bf Epoch B:} During this epoch the available peak optical flux is about a 
factor of $\sim$22 larger than the mean level during the quiescent epoch. An  enhancement in the $\gamma$-ray flux by a factor of  $\sim$13 relative to the quiescent epoch is there 
during the same epoch of the optical flare (see Table \ref{table-2}). From DCF analysis we found the optical and $\gamma$-rays are 
correlated with zero lag (lag = $-$0.25$^{+3.92}_{-6.56}$ days).
In IR too, there are indications of
increased brightness during this epoch, however, the nature of the X-ray state 
of the source
could not be ascertained during this epoch due to the lack of flux measurements. 
Optical polarization measurements were not available during the peak of the optical flare,
but the measurements during the 20 day period centered around the optical flare show
that the source is strongly polarized at the 20\% level, relative to the quiescent
state where the degree of polarization (PD) is around 2\%. Thus in this epoch, there
is an optical flare with a GeV counterpart.

\noindent {\bf Epoch C:} During this period, the source was at its quiescent phase 
in all the wavelengths analyzed in this work. The source was also less polarized during 
this epoch with a PD of around 2\%.

\noindent {\bf Epoch D:} The source has shown an optical flare during this epoch. 
Simultaneous to the optical flare, the IR fluxes too have increased with the peak of 
the IR measurements coinciding with the peak of the optical flare. Around the 
peak of the optical flare the source is detected in the $\gamma$-ray band with 
high significance (TS $>$ 9). Considering points also with TS $<$ 9 during this
epoch, the presence of a $\gamma$-ray flare is evident. This is also confirmed
from the statistical analysis presented in Table \ref{table-2}. From DCF analysis we found a lag of 0.48$^{+6.61}_{-5.48}$ days which is 
consistent with zero.
There is lack of X-ray flux measurements during the peak 
of the optical flare and therefore, the exact X-ray flux state of the source during
the optical flare is not known. Though there are no simultaneous polarization measurements
during the peak of the optical flare, the few measurements available during the 20 days
period of this epoch show that the PD during this epoch is about 30\%, a factor of about 15 
larger relative to the polarization at the quiescent state. Thus in this 
epoch too, there is an optical flare with a GeV counterpart.
\subsubsection{\bf OJ 287}
\begin{figure*}
\hspace*{-3cm}
\includegraphics[width=1.3\textwidth]{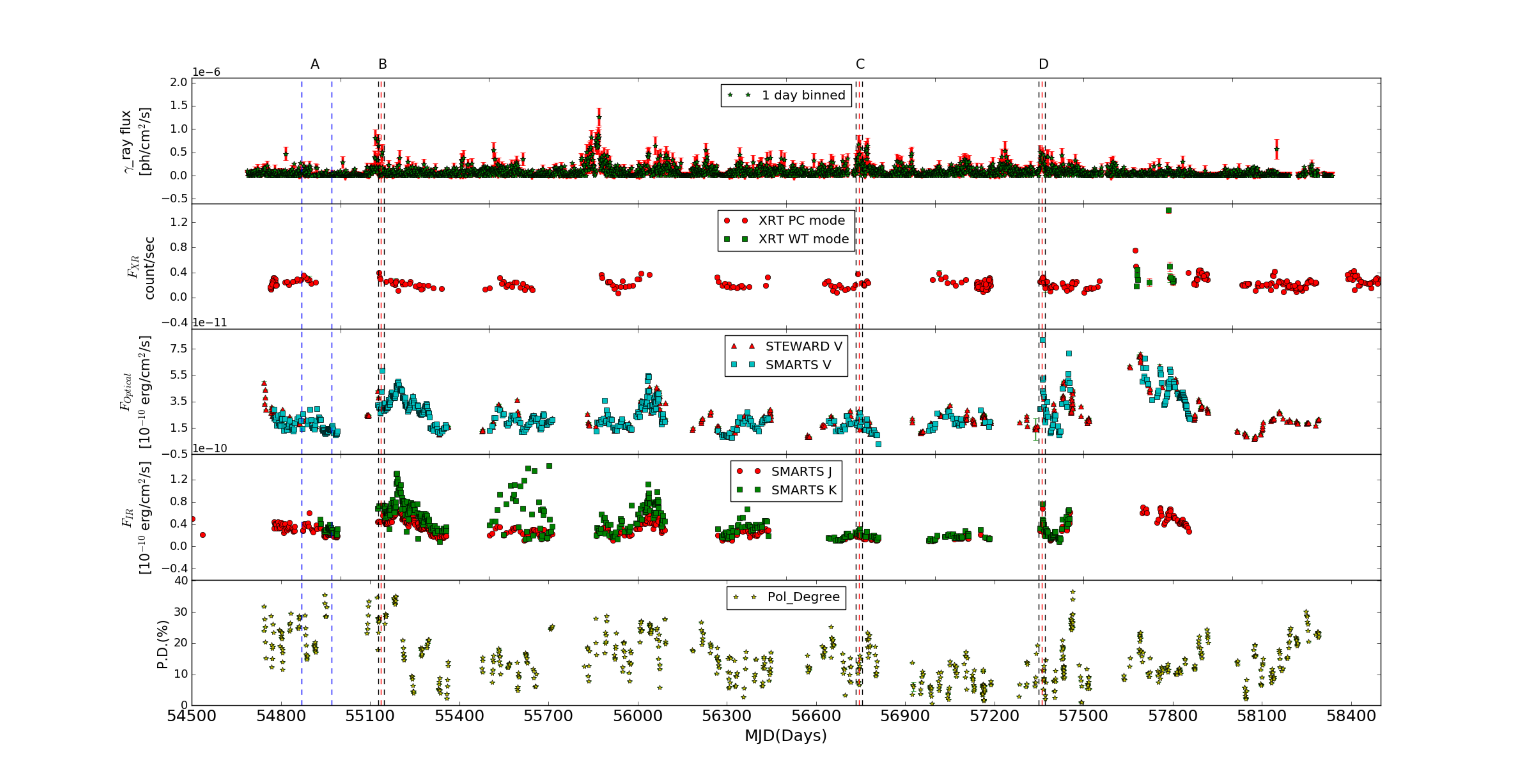}
\caption{Multi-wavelength light curves of the source OJ 287. The other details are similar to that given in the caption to Fig. \ref{figure-1}.} 
\label{figure-3}
\end{figure*}

\begin{figure*}
\begin{center}$
\begin{array}{rr}
\includegraphics[width=75mm,height=85mm]{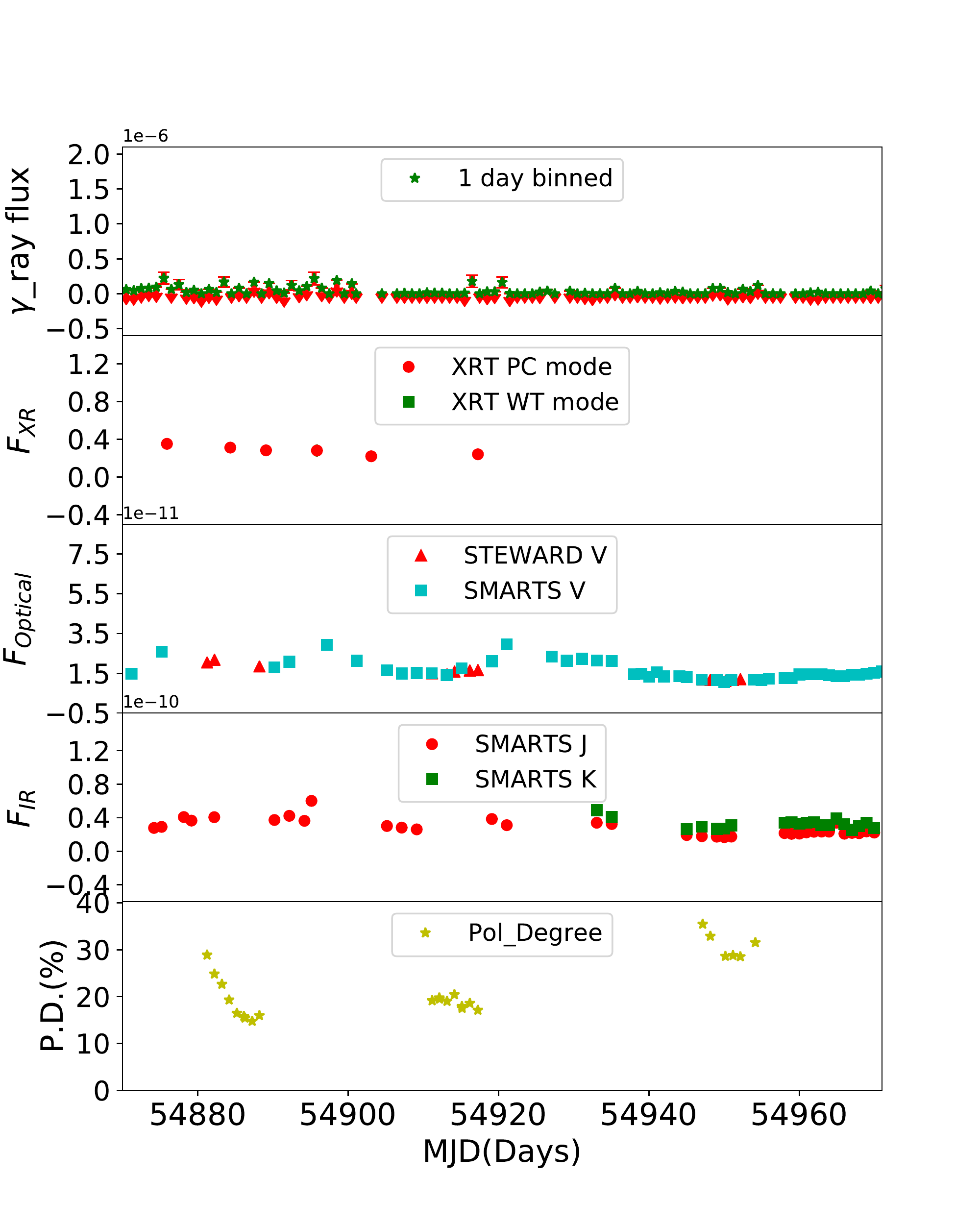}
\includegraphics[width=75mm,height=85mm]{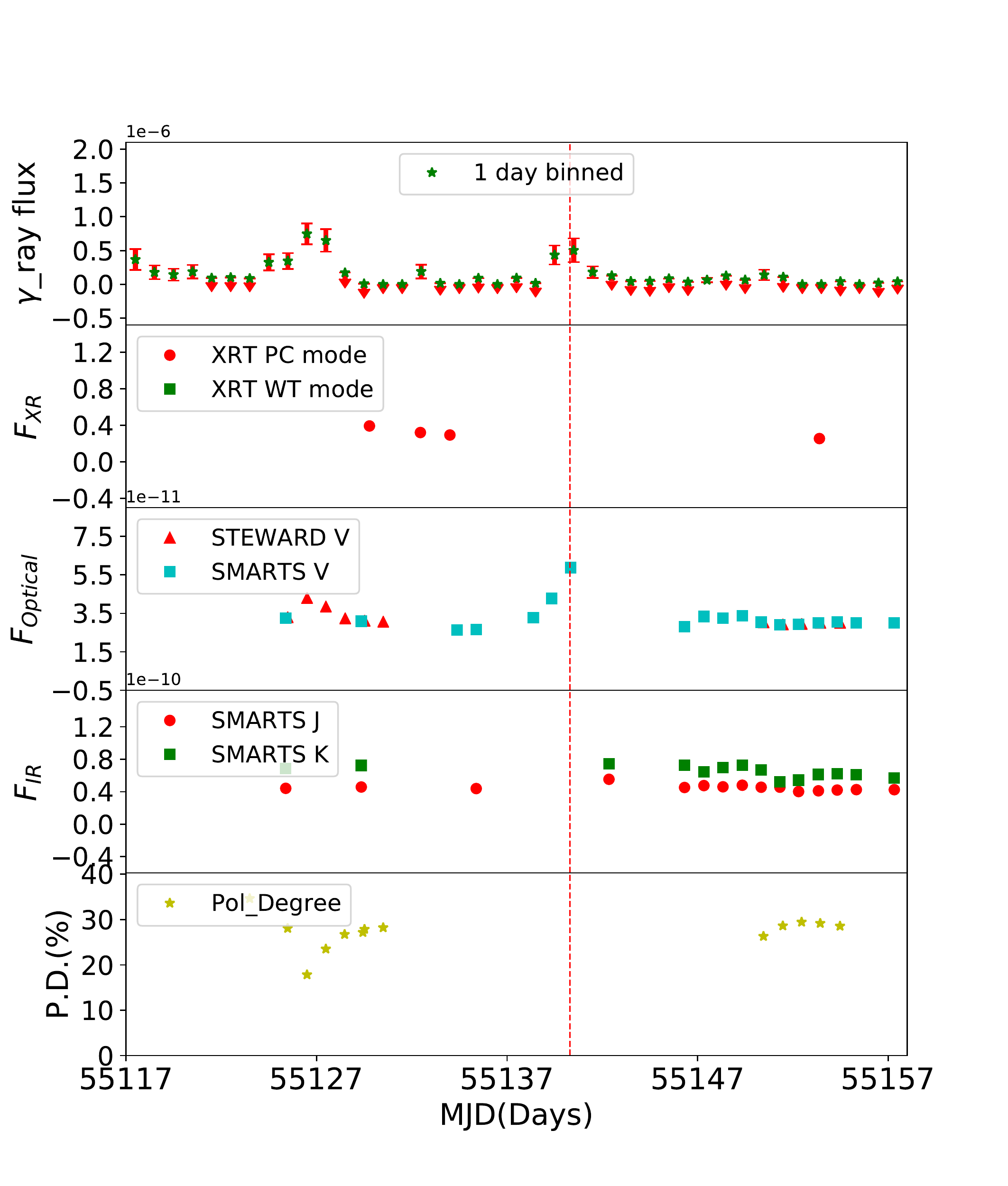}
\end{array}$
\end{center}
\begin{center}$
\begin{array}{rr}
\includegraphics[width=75mm,height=85mm]{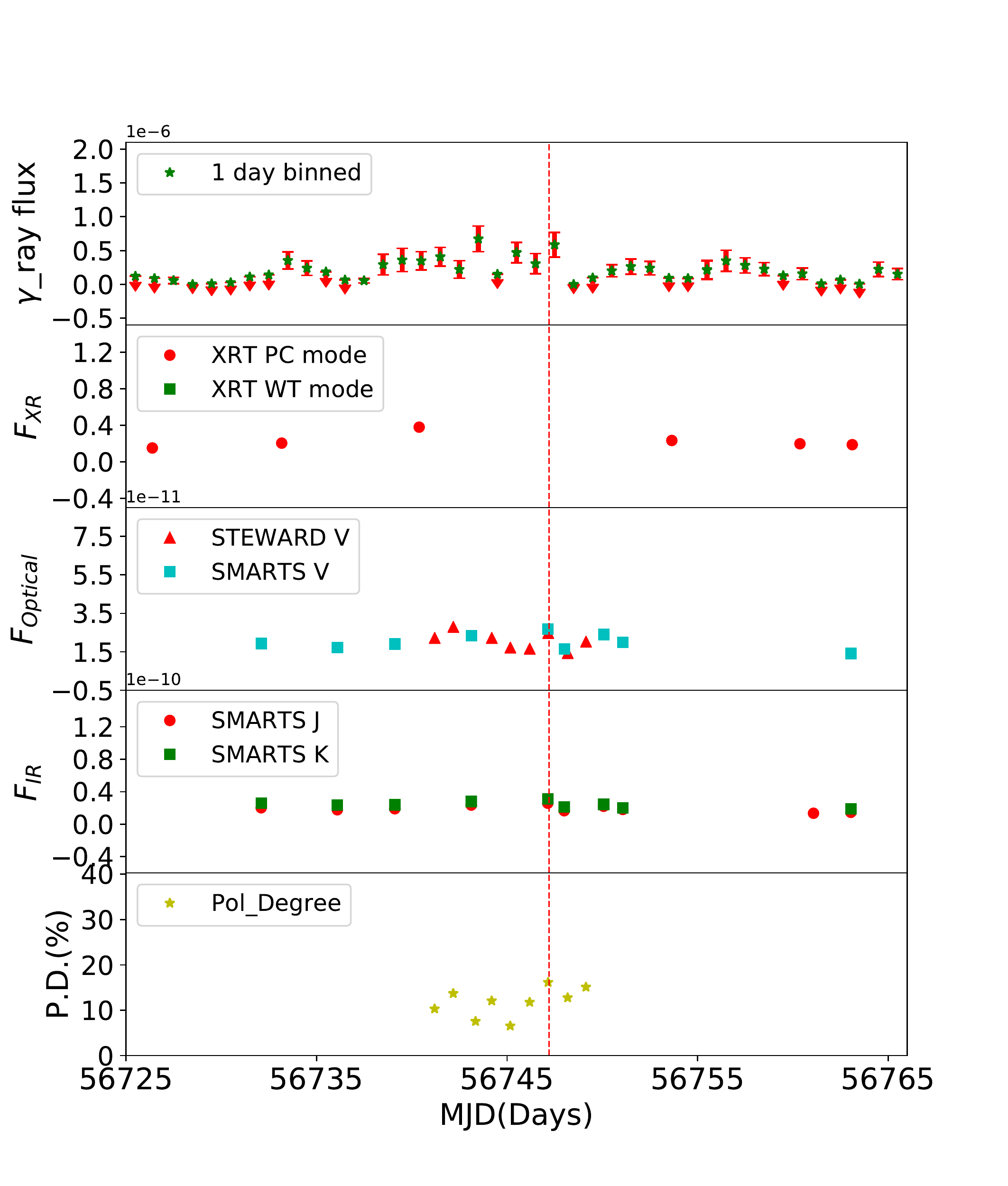}
\includegraphics[width=75mm,height=85mm]{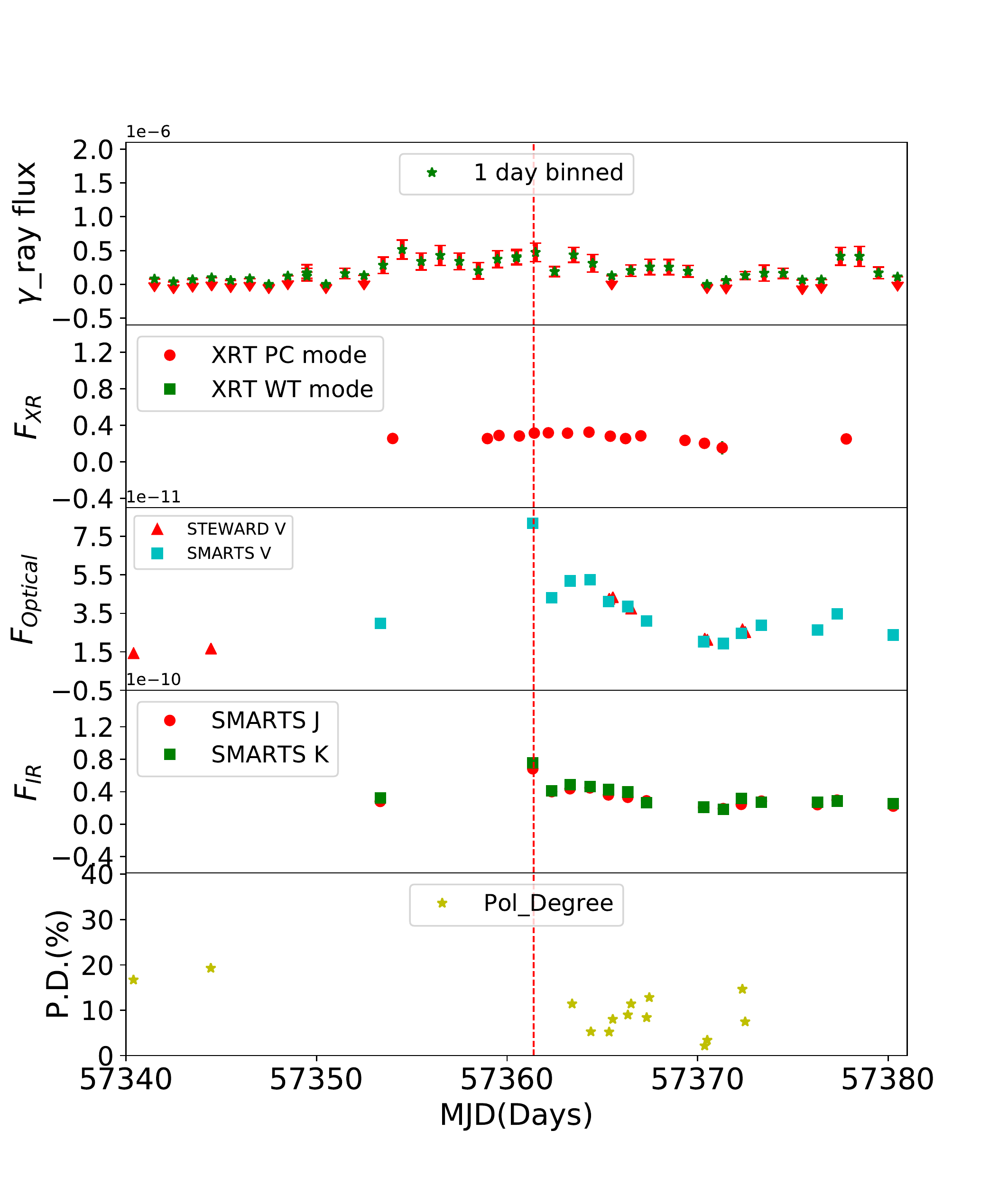}
\end{array}$
\end{center}
\caption{Multi-wavelength light curves for the selected epochs of the source 
OJ 287. The top left and right panels are for epochs A and B, while the bottom left and right panels are for epochs C and D respectively.}
\label{figure-4}
\end{figure*}

We show in Fig. \ref{figure-3} the multi-wavelength light curves of the source, that
include data from {\it Fermi}, {\it Swift}-XRT, SMARTS and the Steward 
Observatory. The source is active all the time in the optical. We identified four epochs
in this source, namely A,B,C and D. The brief details of these four epochs are given in
Table \ref{table-2}. An expanded view of those four epochs, that covers a 
time span of 40 days for the flaring epochs and 100 days for the quiescent epoch is
shown in Fig. \ref{figure-4}. 

\noindent{\bf Epoch A:} During this epoch, the source was at a quiescent state at
all the wavelengths, though some small scale variations are seen in the optical 
light curve. It was below the detection limit (TS $<$ 9) in the $\gamma$-ray band 
for most of the time during this epoch. During this quiescent period of 100 days,
dramatic changes were noticed in the degree of optical polarization. During the beginning
of the epoch, PD decreased from about 30\% to around 15\% in about 20 days,
remained steady at around 20\% in the middle of the epoch and again increased to about
30\% during the end of this epoch. No changes in optical flux were noticed during
the times of polarization variations. 

\noindent{\bf Epoch B:} There is a weak $\gamma$-ray flare in this epoch that 
has a corresponding enhancement in flux in the optical band. The X-ray 
brightness state of the source during the epoch of the weak optical and 
$\gamma$-ray flare is unknown due to the absence of X-ray measurements. Polarization measurements during
the beginning of this epoch indicate that the source has high optical polarization of 
about 20 to 30\%. The ratio of the peak optical and $\gamma$-ray fluxes to their
corresponding mean flux levels during their quiescent 
period are 12.50 $\pm$ 2.37 and 
5.88 $\pm$ 0.02 respectively (see Table \ref{table-2}). Cross-correlation analysis between the optical and $\gamma$-ray 
light curves gave a value of 0.11$^{+1.36}_{-1.21}$ days which is 
consistent with zero. Thus, during this 
epoch optical and $\gamma$-ray flares are correlated.

\noindent{\bf Epoch C:}
During this epoch from visual inspection there is a weak optical flare, 
which again has a correlated
$\gamma$-ray flare that is moderate. Statistical analysis too confirms
this (Table \ref{table-2}). From DCF analysis during this epoch, we found that the optical
and $\gamma$-rays are correlated with zero lag (lag = 2.27$^{+3.45}_{-3.57}$ 
days). The optical polarization also showed a marginal increase during
the peak of the optical and $\gamma$-ray flare. Thus in this epoch there is a correlated optical and $\gamma$-ray flare.

\noindent{\bf Epoch D:}
During this epoch, the source showed a large optical flare. In the IR band too, a flare
is noticed albeit with low amplitude. However, the source was detected in the 
$\gamma$-ray band with high significance (TS $>$ 9) only during the peak 
of the  flare while during most of the time in this epoch, the source
was detected with less significance (TS $<$ 9). Considering all the detections 
along with the statistical analysis (Table \ref{table-2}) indicate the presence
of a $\gamma$-ray flare along with an optical flare. Cross correlation analysis of the optical and $\gamma$-ray light curves
during this epoch shows that the optical and $\gamma$-ray light curves
are correlated with a lag of 2.18$^{+4.79}_{-2.36}$ days. The X-ray brightness during
this epoch was consistent with a constant flux level. Polarization observations
available during the end of the epoch indicates the source to have lower polarization 
compared to the values of PD at epoch B. Thus, during this epoch too we have an optical flare with a $\gamma$-ray counterpart. 

\subsubsection{\bf PKS 2155$-$304}
\begin{figure*}
\hspace*{-3cm}
\includegraphics[width=1.3\textwidth]{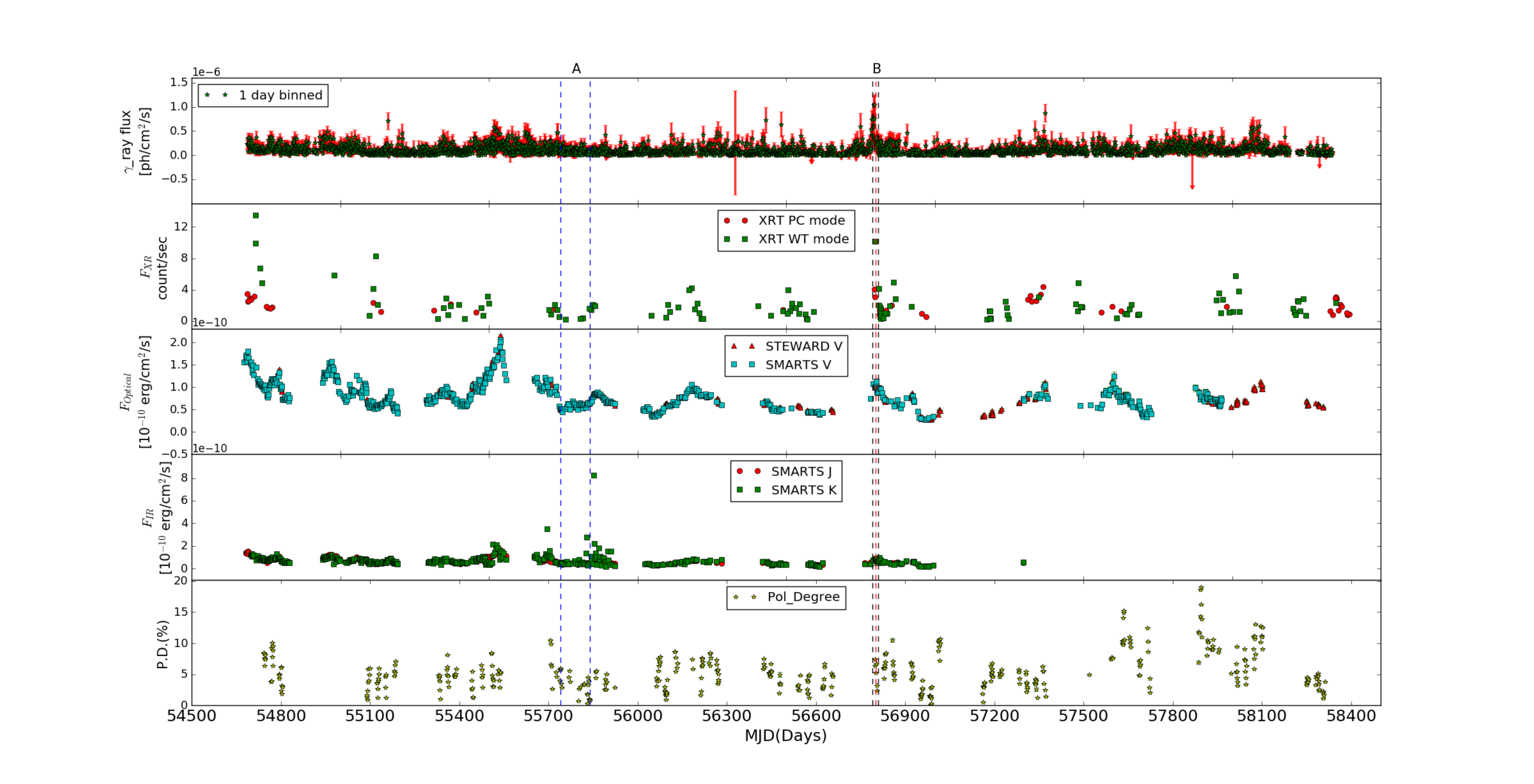}
\caption{Multi-wavelength light curves of the source PKS 2155$-$304. Other details to the figure are similar to that given in the caption to Fig. \ref{figure-1}.}
\label{figure-5}
\end{figure*}

\begin{figure*}
\begin{center}$
\begin{array}{rr}
\includegraphics[width=75mm,height=85mm]{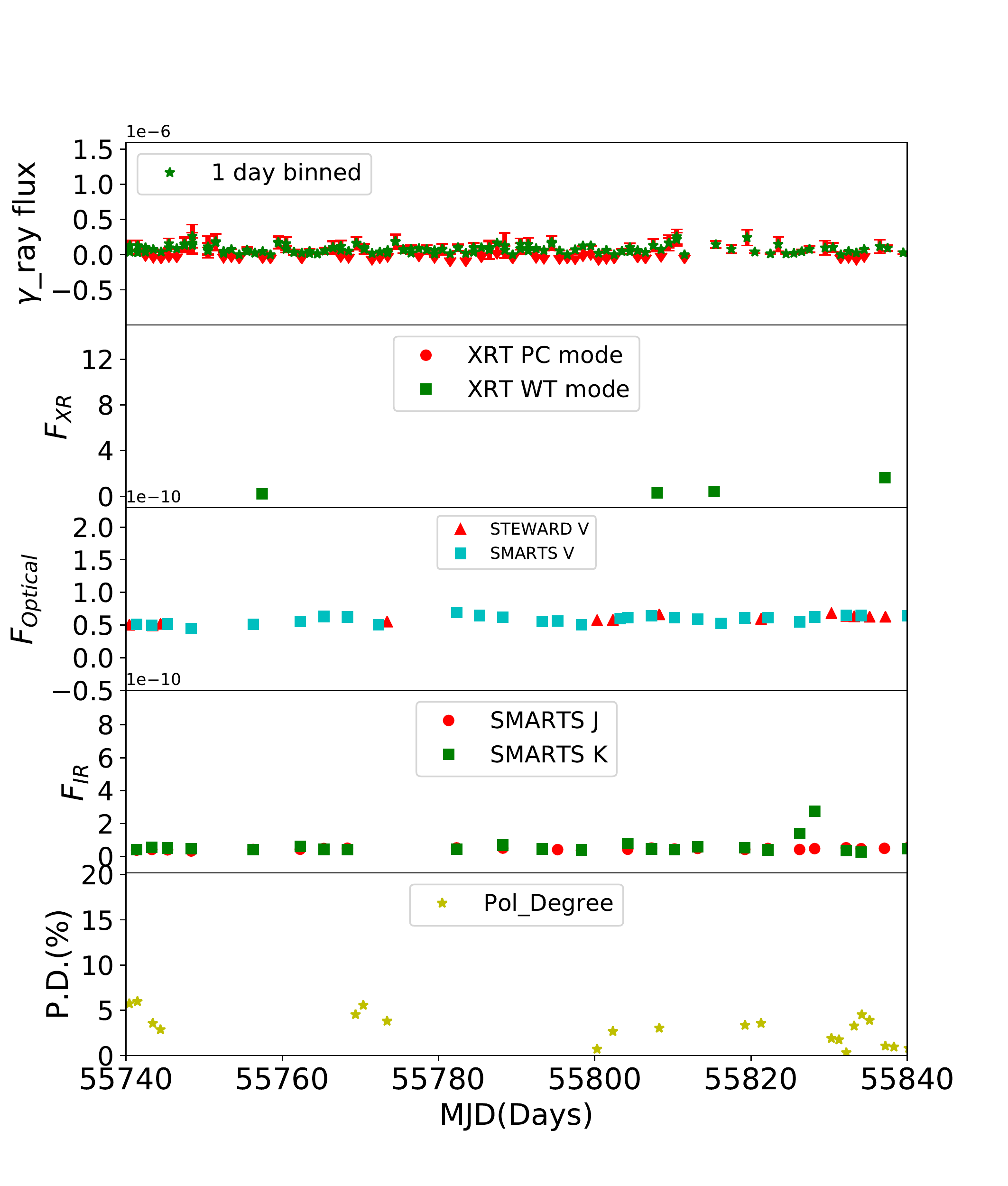}
\includegraphics[width=75mm,height=85mm]{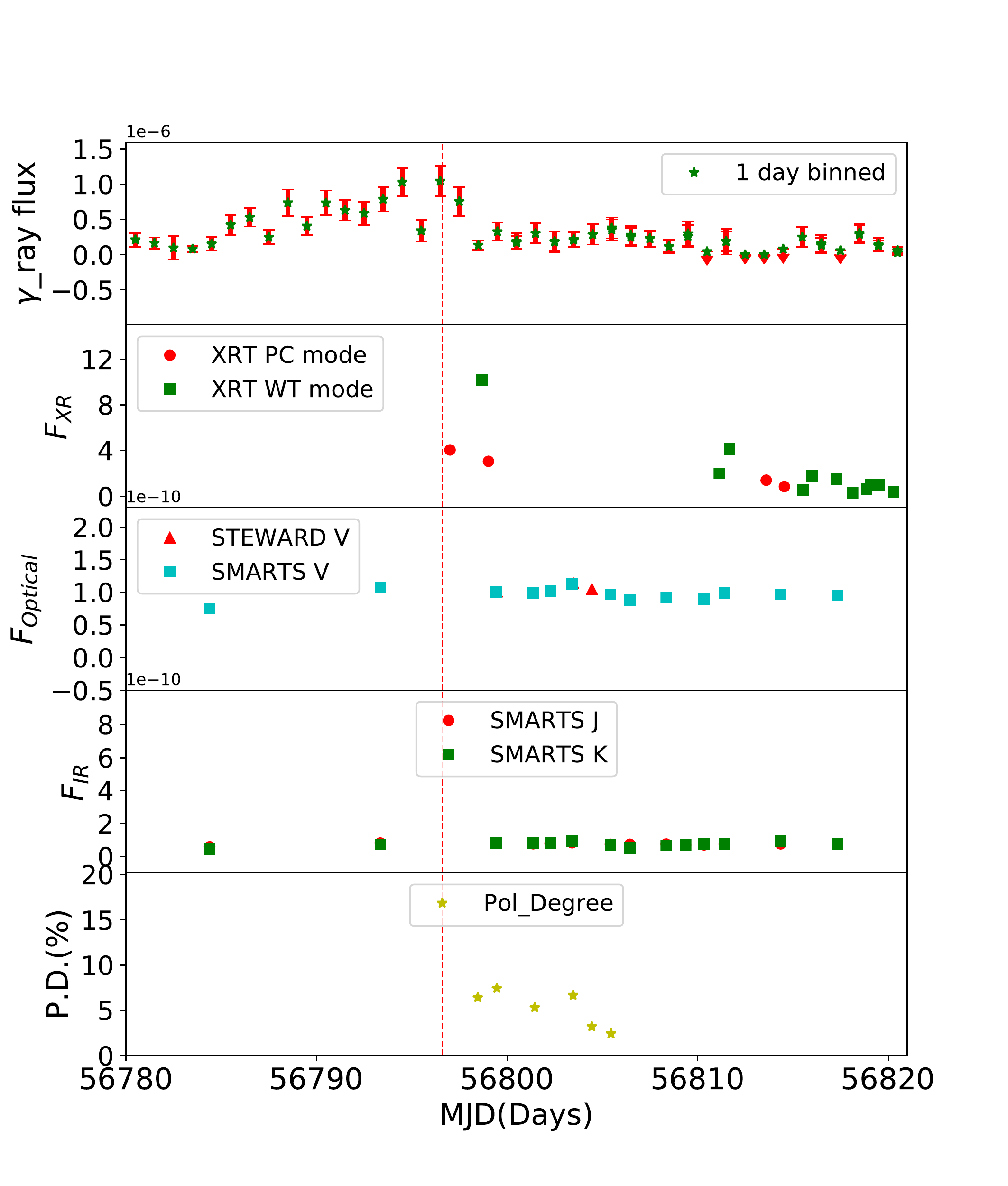}
\end{array}$
\end{center}
\caption{Multi-wavelength light curves for the selected epochs of the source PKS 2155$-$304. The left and right panels are for the epochs A and B respectively.}
\label{figure-6}
\end{figure*}

During the 10 year period analyzed here for flux variations, the source
was found to be variable all the times in the optical band. This is evident in 
Fig. \ref{figure-5} where we show the multi-wavelength light curves of the source. In this
source we identified two epochs namely A and B. The summary of these two epochs are given
in Table \ref{table-2} and an expanded view of these two epochs are shown in
Fig. \ref{figure-6}. For the quiescent period this expanded plot is shown for a 100 day
period, while for epoch B, it is shown for a duration of 40 days. The details of these
two epochs are given below.  

\noindent{\bf Epoch A:} We considered this epoch as the quiescent state of the source. 
During this period, the one day binned $\gamma-$ray light curve was nearly stable. In the
optical and IR bands too, the flux of the source remained stable. Available X-ray flux
measurements during this period too points to the source being weak in X-rays. No changes
were noticed in the optical polarization with the PD remaining nearly constant at a value
of about 3\%. This epoch thus represents the true quiescent period of the source.

\noindent{\bf Epoch B:} The source displayed a strong $\gamma$-ray flare during 
this epoch. Coincident with the $\gamma$-ray flare we noticed a low amplitude 
optical and IR flare. X-ray too showed increased flux during the optical and $\gamma$-ray
flare however the peak of the X-ray does not coincide with the peak of the $\gamma$-ray flare. Although the overall optical brightness state of the 
source is larger compared to the other periods, there is lack of optical 
data at the peak of the $\gamma$-ray flare. However visual inspection of 
Fig. \ref{figure-5}, as well as  statistical analysis (Table \ref{table-2}) 
indicate a $\gamma$-ray flare correlated with an  
optical flare. Cross correlation analysis indicate that the optical
and $\gamma$-ray light curve during this epoch are correlated with lag of 6.08$^{+1.74}_{-7.89}$. Optical polarization was also higher during this epoch relative 
to the quiescent
epoch A. Thus during this epoch we observed correlated flux variations in IR, optical, 
and $\gamma$-rays. 

\subsection{Spectral Variations}
\begin{figure*}
\hspace*{-0.9cm}\includegraphics[scale=0.60]{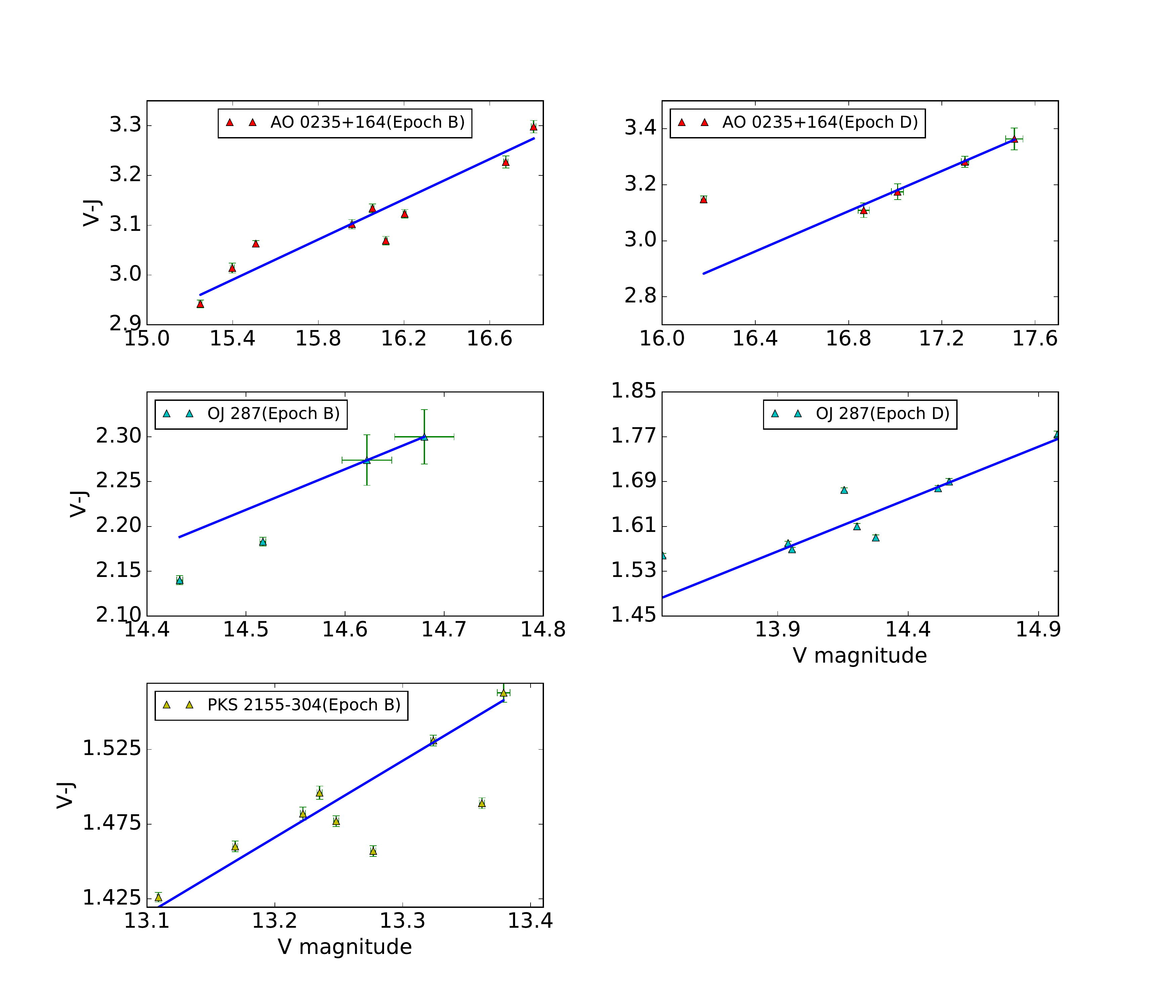}
\caption{Colour-magnitude diagram. The top left and right panels are for AO 0235+164 epochs B and D respectively. The middle panels are for the epoch B (left panel) and epoch D (right panel) of OJ 287 and the bottom panel is for the epoch B of PKS 2155$-$304.}
\label{figure-7}
\end{figure*}

The flux variations in the optical and IR bands shown by blazars are accompanied
by spectral/colour variations. To investigate the spectral variability 
characteristics of the BL Lacs studied in work, we generated color (V-J)-magnitude (V) diagrams of all the 
selected epochs for the three BL Lac objects and checked for the correlation 
of V-J band color against the V-band brightness. To characterize spectral variations,
we carried out linear least squares fit to the colours and magnitudes taking into
account the errors in both of them. We considered a source to show colour 
variation if the Spearman rank correlation coefficient is $>$ 0.5 or $<$ $-$0.5 and the 
probability of no correlation is less than 0.05, so that the claimed correlation is
significant at the 95\% level. In the source A0 0235+164, we found "bluer when
brighter (BWB)" trend for the 
epochs A and B. In OJ 287, we found BWB trend at epochs B and D, while for 
PKS 2155$-$304 we found BWB trend for the lone epoch B. Thus in all the sources for
the epochs where a statistically significant colour magnitude relation could be established, we found a BWB trend. The colour magnitude relation for all the sources that satisfies
the statistical criteria outlined above along with the
linear least squares fit to the data are shown in Fig. \ref{figure-7}. 

Studies on optical-IR colour variations in blazars generally point to FSRQs showing
a redder colour with increasing brightness (RWB; \citealt{2019ApJ...887..185S}) 
and BL Lacs showing a bluer colour with increasing brightness 
(BWB; \citealt{2019MNRAS.484.5633G}). Recent studies on the colour variations in 
blazars, show that in FSRQs, both BWB as well as RWB trends are seen 
\citep{2019MNRAS.486.1781R,2020MNRAS.tmp.2538S,2020MNRAS.tmp.2558R}. 
In the objects studied here, whenever statistically significant colour variations 
were observed, we found a BWB trend (see Fig. \ref{figure-7}). Such a BWB trend 
could happen because of changes in the Doppler factor (\citealt{2004A&A...421..103V,2007A&A...470..857P}). 
In can also happen due to increase in the amplitude of variations at shorter 
wavelengths \citep{2009MNRAS.399.1357S} which in the one zone leptonic scenario 
can happen due to the injection of fresh electrons that have an energy 
distribution that is harder than older softer elections 
\citep{1995A&A...295..613M}. Though our analysis points to BL Lacs showing a BWB 
trend, it is unlikely they do not show a RWB trend. For example based on an 
analysis of about 10 years of optical - IR data on AO 0235+164, an LSP BL Lac, 
\cite{2020MNRAS.tmp.2538S} noticed BWB trend upto certain optical V-band 
brightness, beyond which the source showed a RWB behaviour.

\subsection{$\gamma$-ray spectrum}

\begin{figure*}
\begin{center}$
\begin{array}{lll}
\includegraphics[width=90mm,height=70mm]{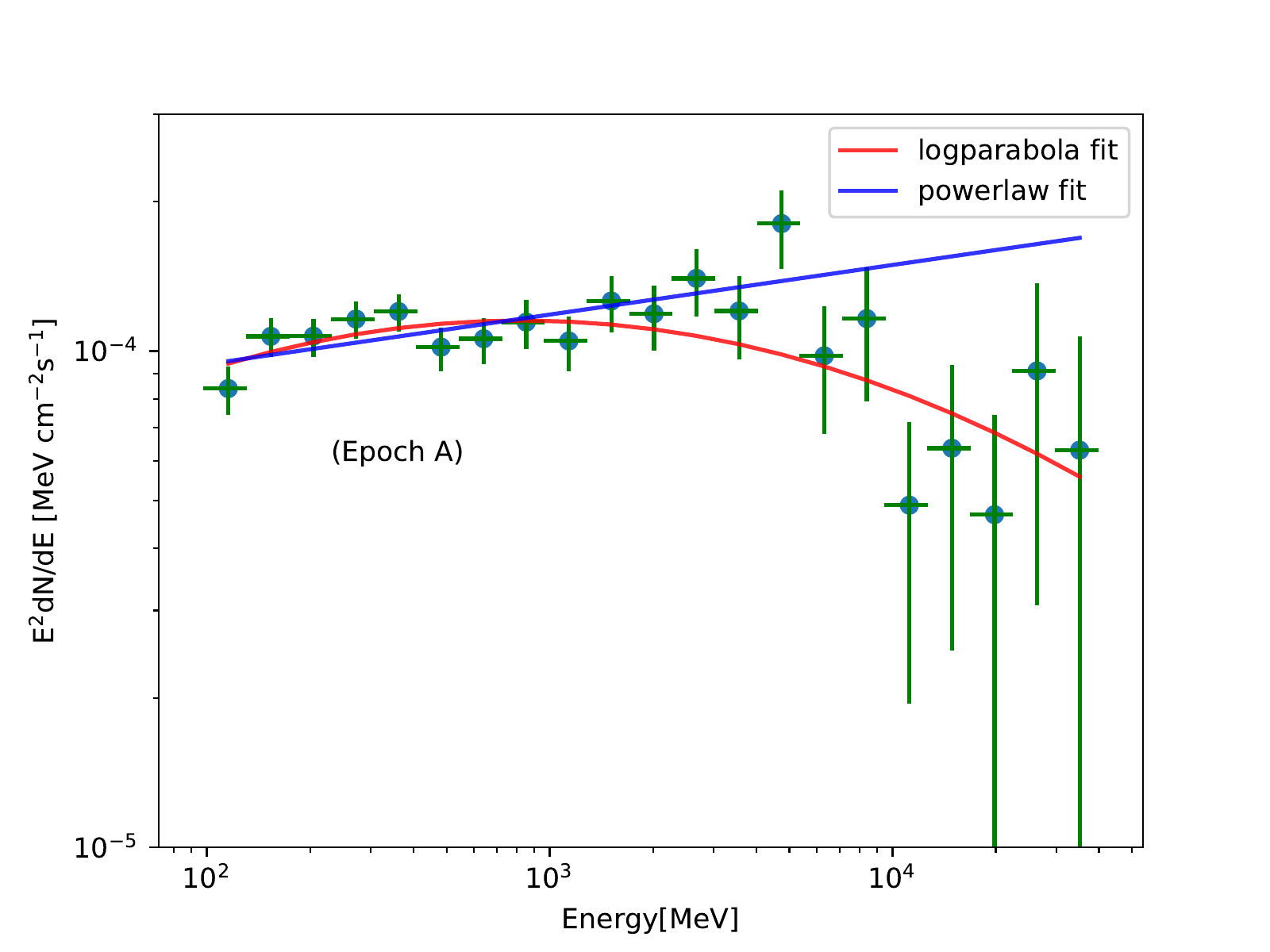}&
\includegraphics[width=90mm,height=70mm]{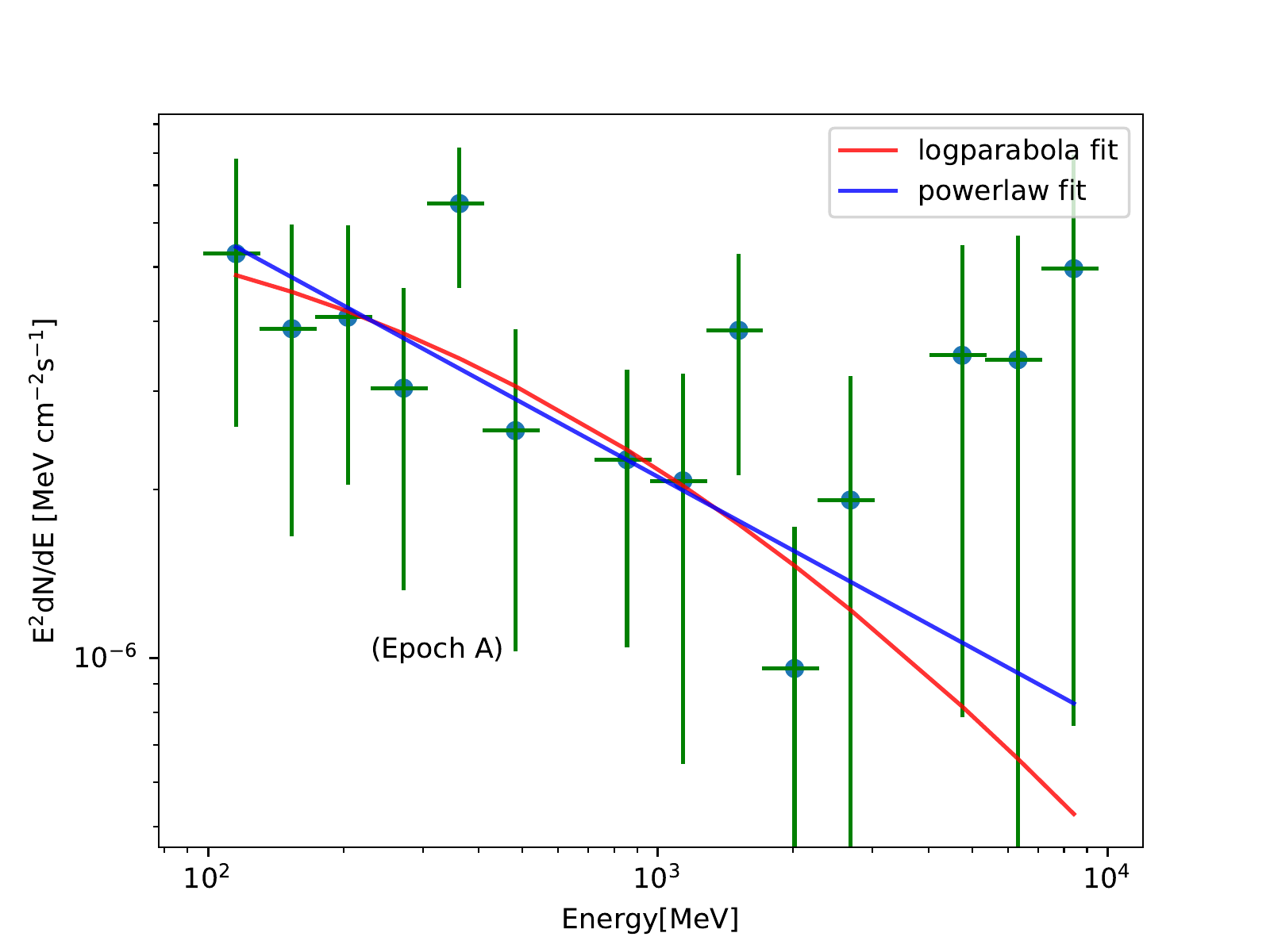}&
\end{array}$
\end{center}
\vspace*{-0.6cm}\caption{Observed and model fits to the $\gamma$-ray spectra.
Left: Epoch A of AO 0235+164 well fit by the LP model and Right: Epoch A of 
OJ 287 well described by the PL model.} 
\label{figure-8}
\end{figure*}

In order to explore the behaviour of $\gamma$-ray spectra for different epochs and to discern the intrinsic distribution of electrons in the jet of BL Lacs, we performed the fitting of the $\gamma$-ray spectra with two models namely power law (PL) and Log parabola (LP). For $\gamma$-ray spectral analysis, the data 
were averaged over a duration of 100-days in the case of quiescent periods and
20 days during other periods. The PL and LP \citep{2012ApJS..199...31N} models are defined as follows :-

\begin{equation}
dN(E)/dE=N_{\circ}(E/E_{\circ})^{-\Gamma}
\end{equation}
and
\begin{equation}
dN(E)/dE=N_{\circ}(E/E_{\circ})^{-\alpha-\beta ln(E/E_{\circ})}
\end{equation}
where, dN/dE is the number of photons in cm$^{-2}$ s$^{-1}$ MeV$^{-1}$, $\Gamma$ is photon index for PL fitting, $\alpha$ is the photon index at $E_{\circ}$, $\beta$ is the parameter that defines the curvature around the peak, E is the energy of the $\gamma$-ray photon, N$_{\circ}$ is the normalization and E$_{\circ}$ is the scaling factor. We used {\tt gtlike}, maximum likelihood estimator to verify 
the model that fits the $\gamma$-ray spectra well. We computed the test 
statistics to check for the presence of curvature \citep{2012ApJS..199...31N} as $TS_{curve}$ = 2(log $L_{LP}$ - log $L_{PL}$). 
For the existence of a statistically significant curvature in the $\gamma$-ray spectra we used the threshold $TS_{ curve}$ > 16. Two sample spectral fits are shown in Fig. \ref{figure-8}, one
in which the spectrum is well described by the LP model and the other in which the spectrum is well described
by the PL model.The large error bars in the $\gamma$-ray spectra are due to poor 
photon statistics. The results of the model fitting are given in 
Table \ref{table:table-3}. For most of the epochs, the $\gamma$-ray spectra are 
well fit by a LP model. 
 
\begin{table*}
\caption{Details of the PL and LP model fits for the selected epochs of the sources AO 0235+164, OJ 287 and PKS 2155$-$304. Here the $\gamma$-ray flux value is in units of $10^{-7}$ph $cm^{-2}$ $s^{-1}$. The value of $\Gamma$, $\alpha$ and $\beta$ mentioned here are obtained by fits to the data which matches with the values returned by fermipy.}
%%\resizebox{\textwidth}{!}{\begin{tabular}{ccccccccccc}
{\begin{tabular}{cccrrccrrrr}
     \hline
	&     \multicolumn{4}{c}{PL} & \multicolumn{5}{c}{LP} &\\ 
     \cline{2-5} \cline{6-10}
%     %\cmidrule(lr){6-12}
         Epochs  & $\Gamma$ & Flux  & TS & $-$Log L  & $\alpha$ & $\beta$ & 
    Flux & TS & $-$Log L & TS$_{curve}$\\
%     %\cmidrule(lr){2-5} 
%     %\cline{2-6}
     \hline
AO 0235+164 &             &              &         &         &                &               &                 &         &          &        \\
  A          &-1.90$\pm$0.02&3.90$\pm$0.13& 3874.64 & 31241.99 & 1.94$\pm$0.03 & 0.05$\pm$ 0.02 &  9.64$\pm$0.12 &  4413.69 & 30888.09 & 707.8 \\
  B          &-1.95$\pm$0.01&3.60$\pm$0.05& 2617.2 &  27655.91 & 1.98$\pm$0.03 & 0.08$\pm$ 0.03 & 8.64$\pm$0.12 & 2982.15 &  27330.24 & 651.36 \\
  C          &-2.40$\pm$0.16&0.19$\pm$0.11& 71.97 & 69867.23 & 2.40$\pm$0.18 & 0.06$\pm$ 0.06 &  0.92$\pm$0.12 &   133.52 & 69843.77 &  46.94 \\
  D          &-1.91$\pm$0.16&1.03$\pm$0.12& 87.59 & 21402.72 & 1.91$\pm$0.18 & 0.07$\pm$ 0.00 &  1.42$\pm$0.05 &   95.85 &  21422.97 & -40.50  \\
              \hline  
OJ 287       &             &              &         &         &                &               &                 &         &          &        \\
 A           &-2.44$\pm$0.13&0.43$\pm$0.05& 104.512 & 57896.37 & 2.48$\pm$0.16 & 0.07$\pm$ 0.17 &  0.40$\pm$0.02 &  98.39 & 57897.17 & -0.16 \\
 B           &-2.51$\pm$ 0.16 & 0.87$\pm$0.19& 109.77 & 15636.63 & 2.95$\pm$0.41 & 0.30$\pm$ 0.27 & 1.47$\pm$0.09 & 143.06 & 15634.43 & 4.40 \\
 C          &-1.88$\pm$0.11&1.77$\pm$0.37&  205.64 & 8613.66 & 1.87$\pm$0.15 & 0.07$\pm$ 0.02 &  2.46$\pm$0.20 &   242.103 & 8601.86 & 23.60 \\
  D          &-2.72$\pm$0.14&1.97$\pm$0.24&  184.95 & 9139.01 & 2.88$\pm$0.40 & 0.09$\pm$ 0.23 &  3.01$\pm$2.32&  254.03 & 9129.45 &  19.12 \\
     \hline     
PKS 2155-304       &             &              &         &         &                &               &                 &         &          &        \\
 A           &-1.78$\pm$0.04&0.41$\pm$0.04& 831.04 & 51370.55 & 1.84$\pm$0.05 & 0.09$\pm$ 0.04 &  0.85$\pm$0.06 &  934.67 & 51344.24 & 52.62 \\
 B           &-1.62$\pm$ 0.04 & 4.09$\pm$0.40& 1541.92 &  14844.11 & 1.83$\pm$0.06 & 0.02$\pm$ 0.03 & 4.97$\pm$0.54 & 1621.77 &  14822.00 & 44.22 \\
     \hline       
     \end{tabular}}
%%\footnotetext[1]{$\gamma$-ray flux value in units of $10^{-6}$ph $cm^{-2}$ $s^{-1}$}
	 \label{table:table-3}
\end{table*}

\subsection{Spectral energy distribution modelling}
For all the sources, when flares were identified visually and then
confirmed statistically, we found close correlation between optical and
$\gamma$-ray flux variations.

\begin{figure*}
\vspace*{-0.5cm}
\vbox{
\includegraphics[scale=0.4, angle=270]{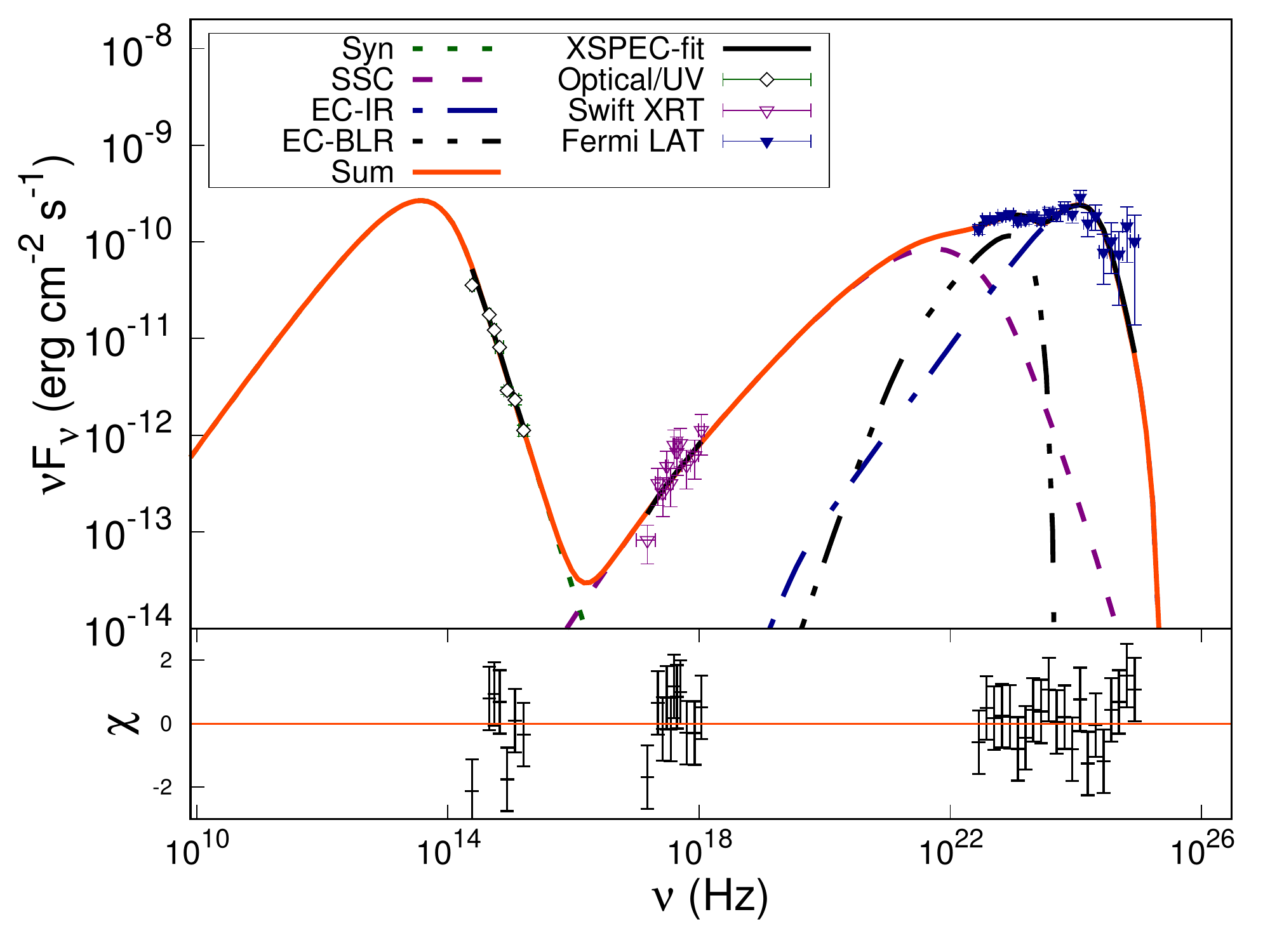}
\includegraphics[scale=0.4, angle=270]{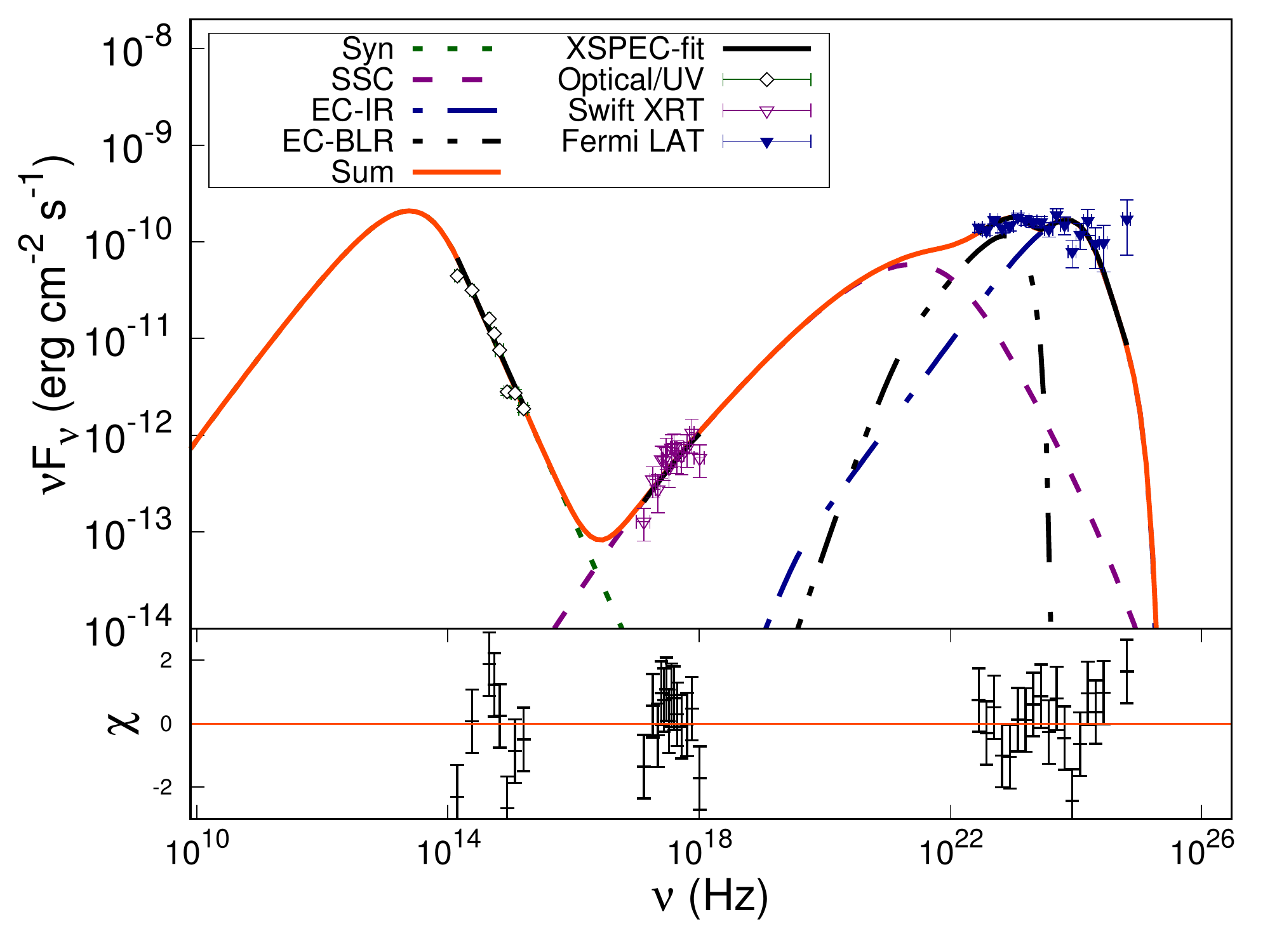}
\includegraphics[scale=0.4, angle=270]{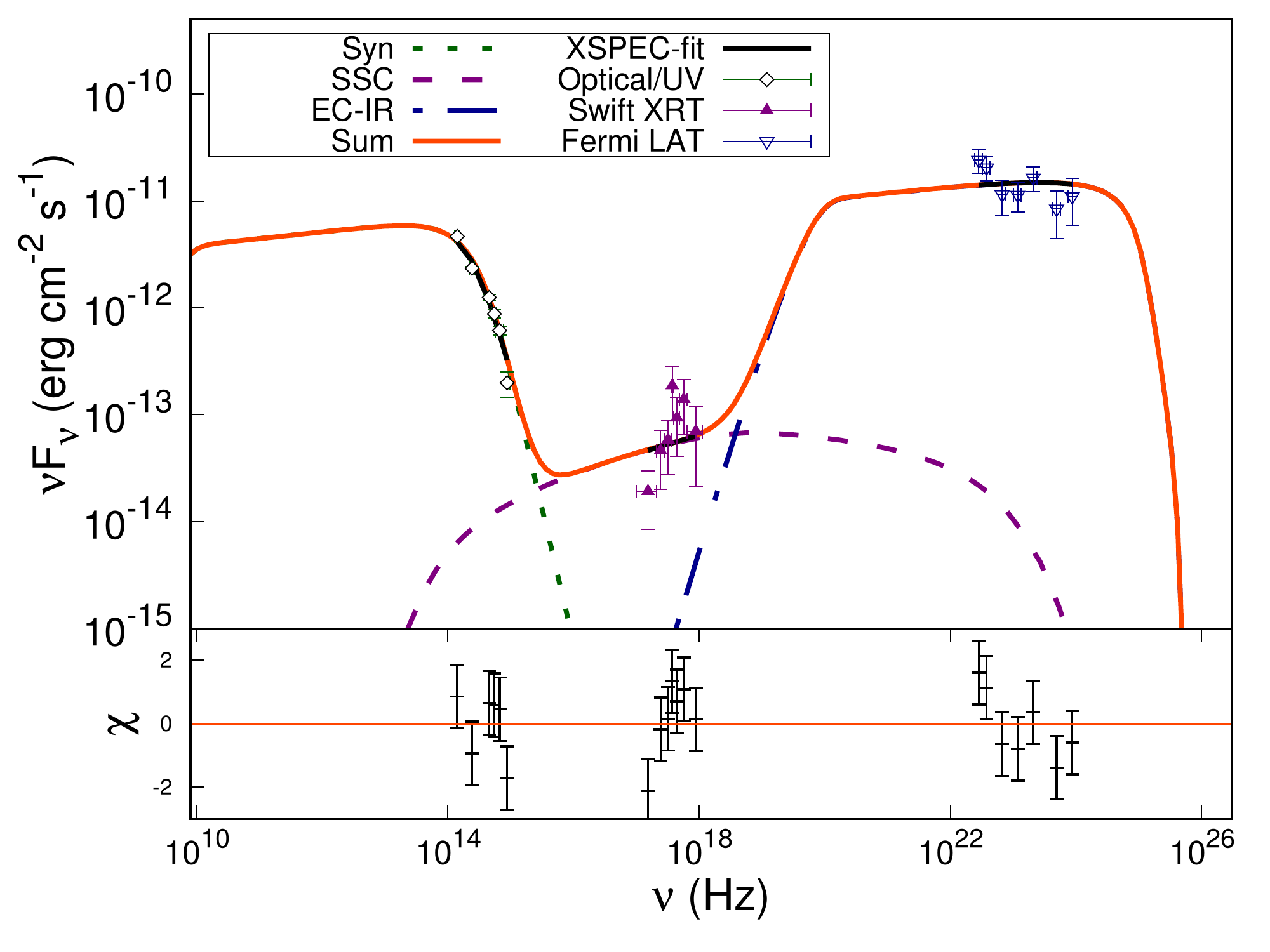}
\includegraphics[scale=0.4, angle=270]{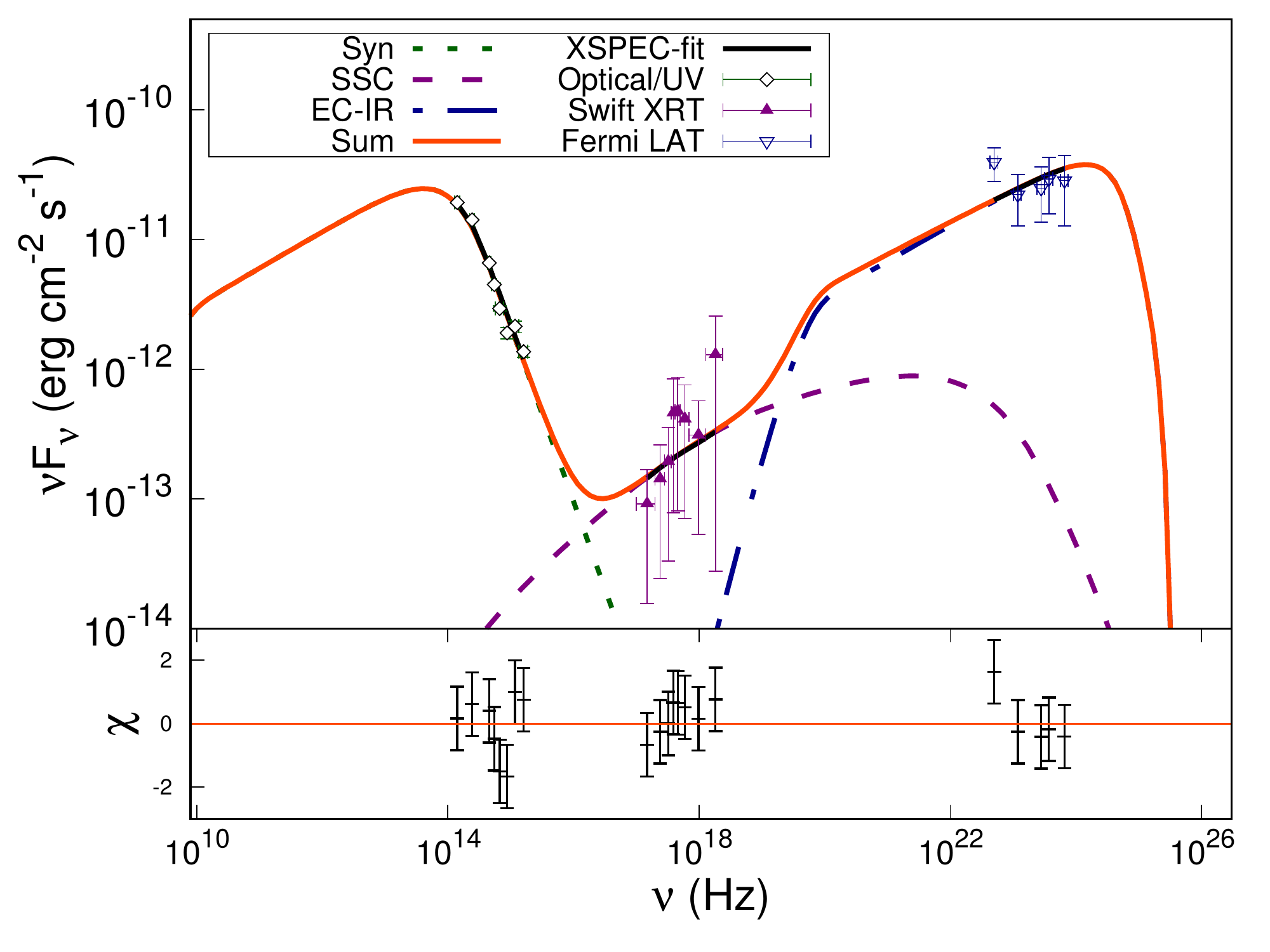}
     }
\vspace*{0.5cm}\caption{One zone leptonic model fits to the broad band SED for epochs A (left-top), B (right-top), C (bottom-left) and D (bottom-right) for the source AO 0235+164. 
%In the figures, the green line is the synchrotron emission, the yellow and red lines are the SSC and EC components respectively. The cyan line is the sum of all the components. 
The second panel in the figures show the residuals, which is estimated from the observed data fitting in XSPEC.}
\label{figure-9}
\end{figure*}

\begin{figure*}
\vspace*{-0.5cm}
\vbox{
\includegraphics[scale=0.4, angle=270]{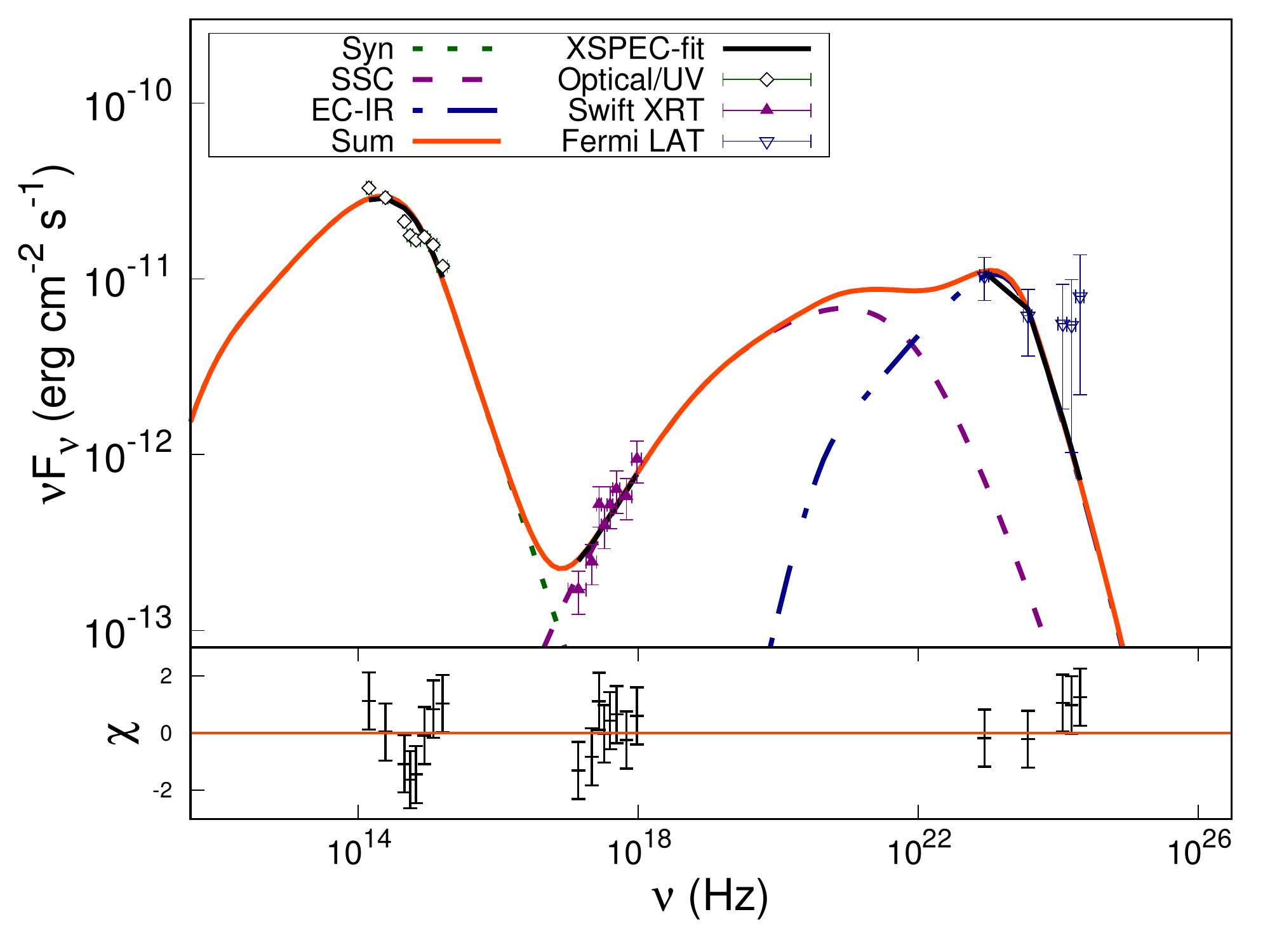}
\includegraphics[scale=0.4, angle=270]{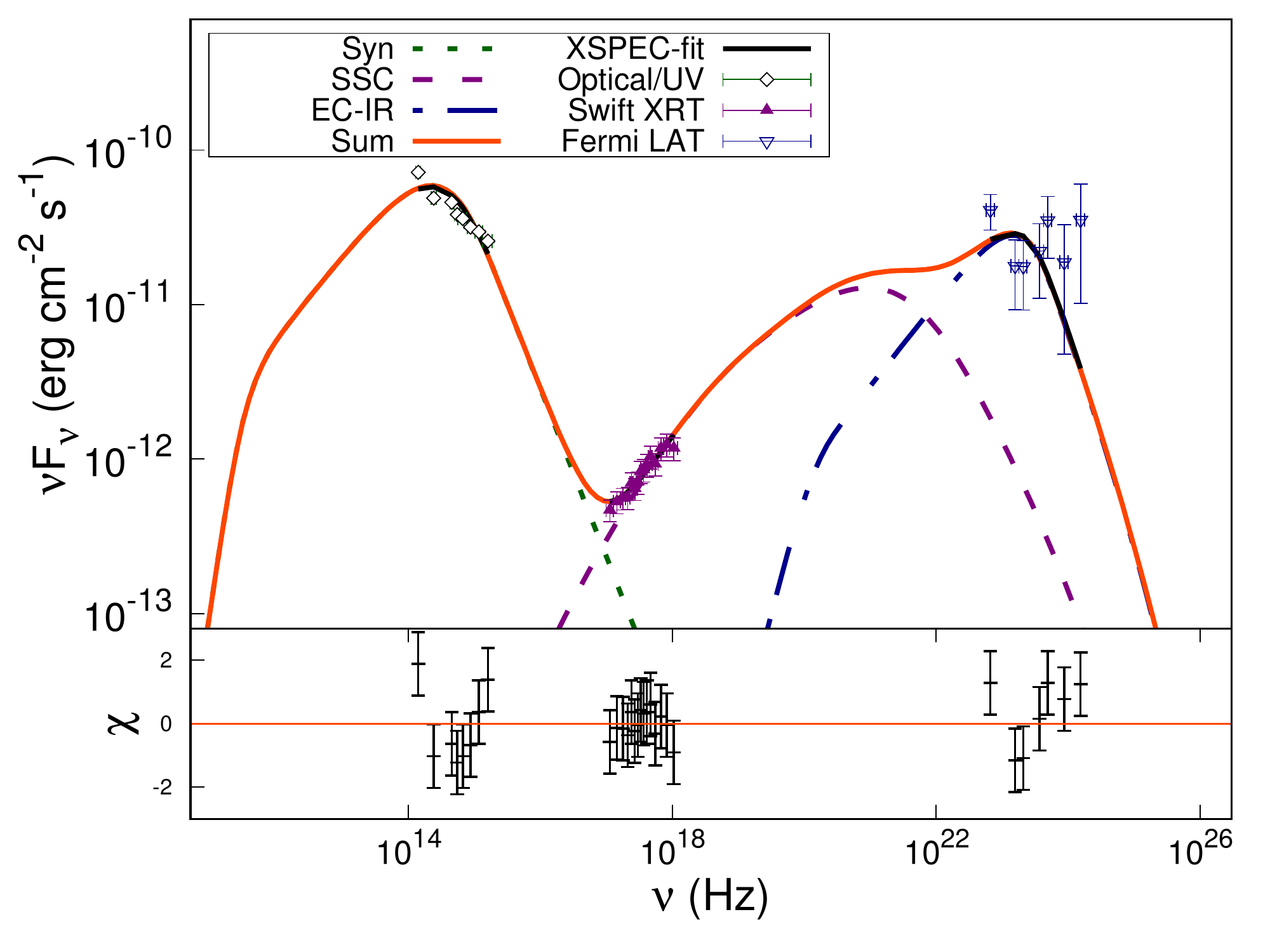}
\includegraphics[scale=0.4, angle=270]{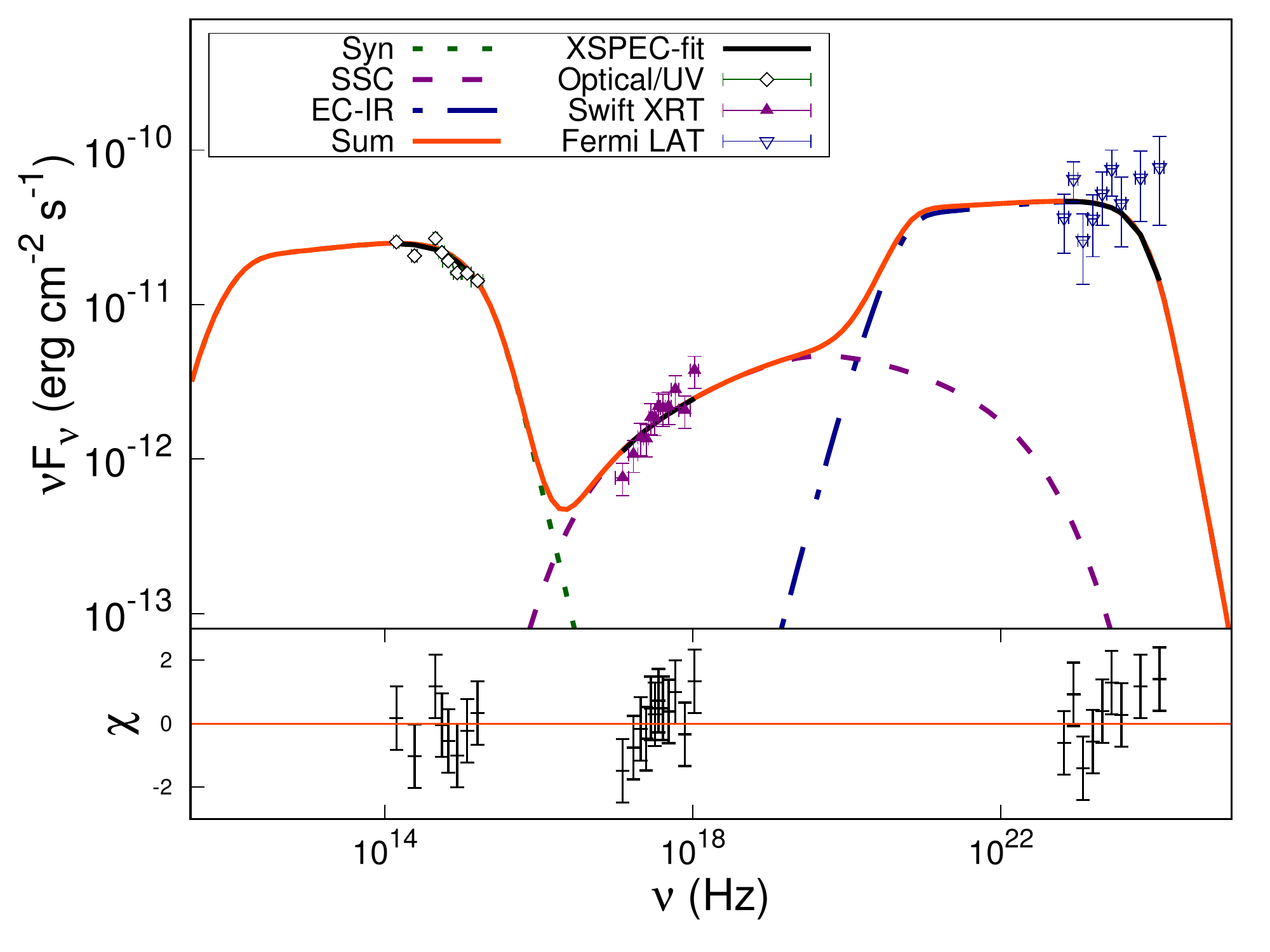}
\includegraphics[scale=0.4, angle=270]{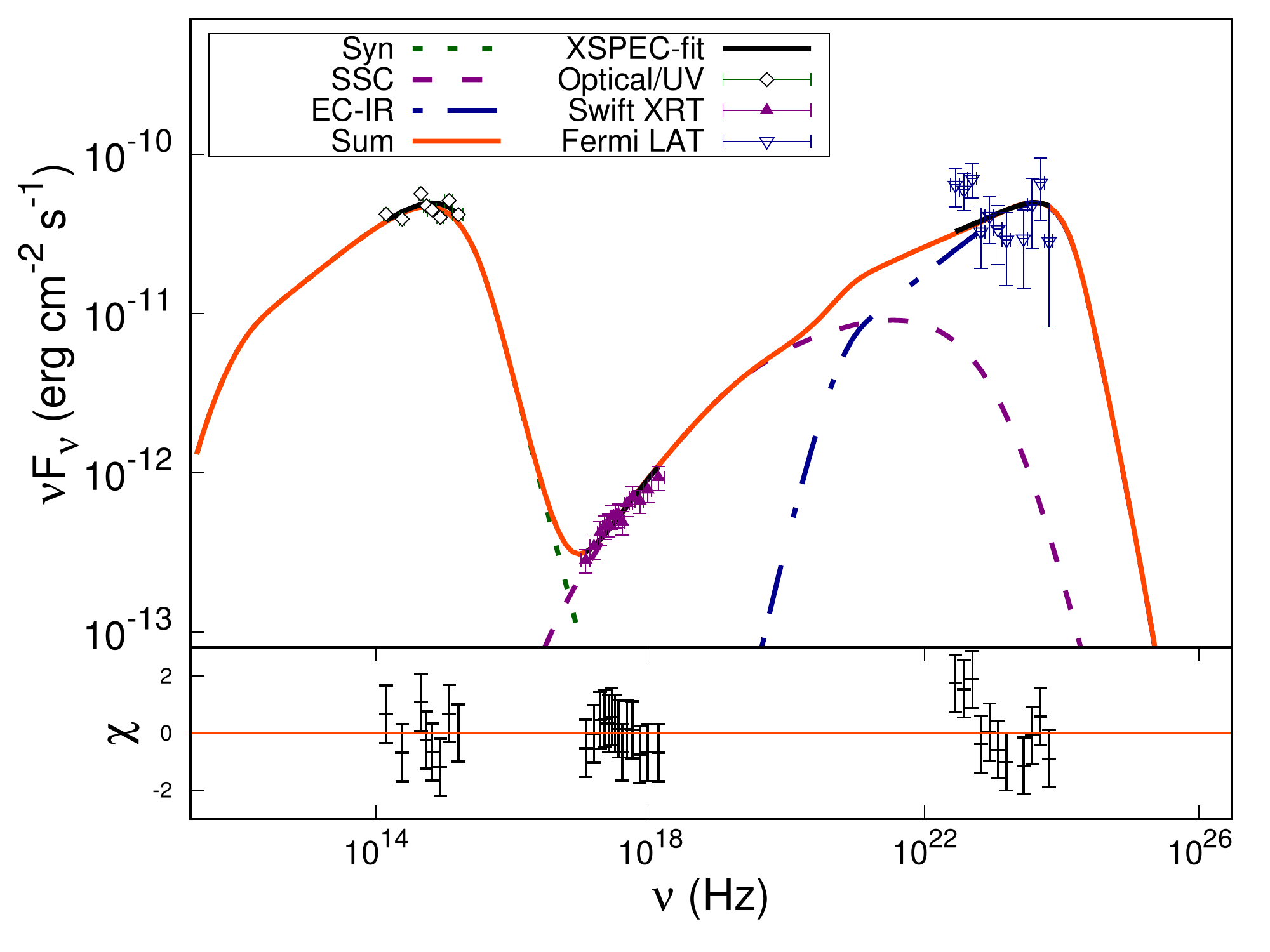}
     }\vspace*{0.5cm}\caption{One zone leptonic model fits to the broad band SED for epochs A (left-top), B (right-top), C (bottom-left) and D (bottom-right) for the source OJ 287. The details to the figures are the same as those
given in Figure \ref{figure-9}.} 
\label{figure-10}
\end{figure*}

\begin{figure*}
\vspace*{-0.5cm}
\vbox{
\includegraphics[scale=0.4, angle=270]{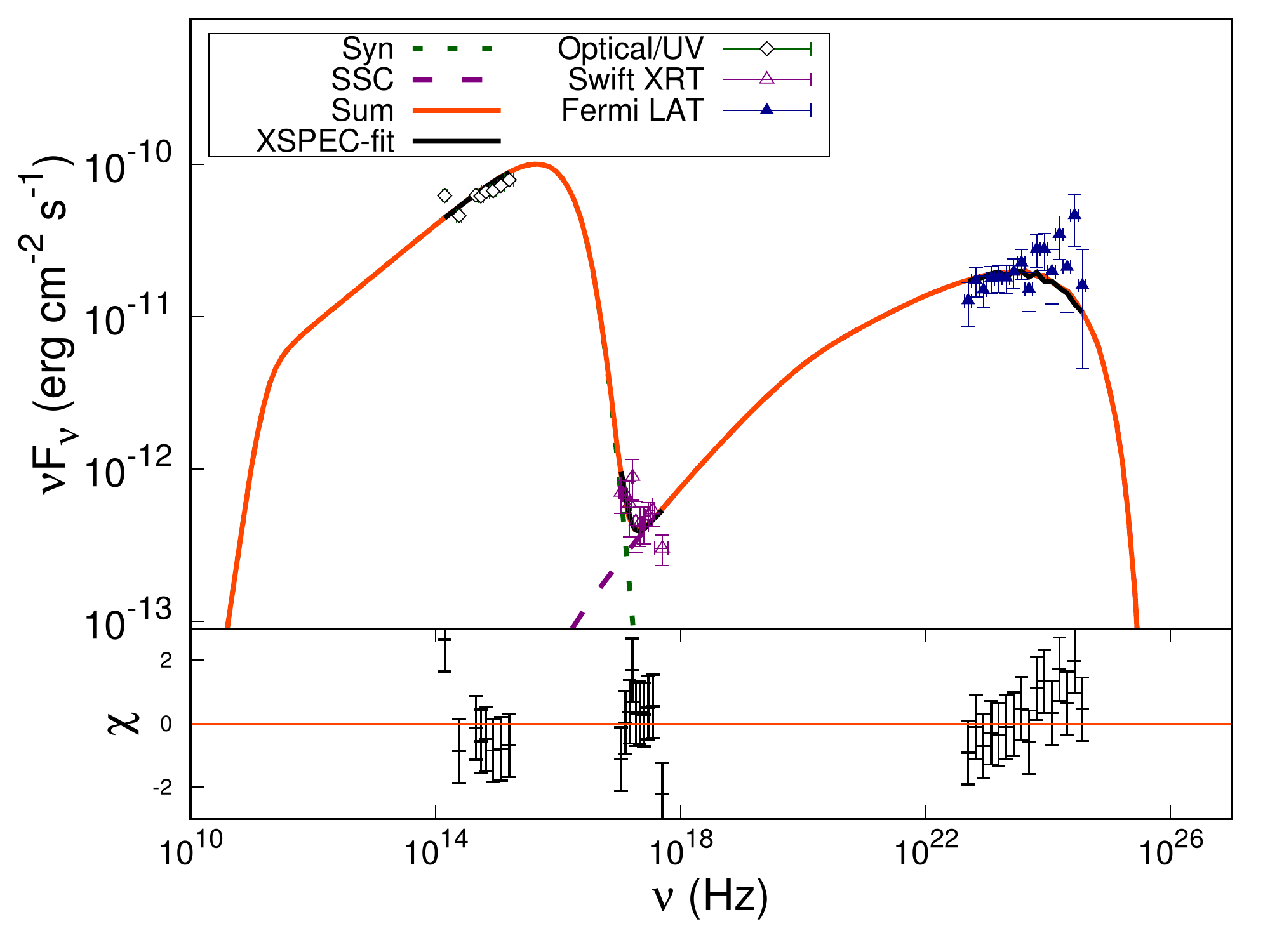}
\includegraphics[scale=0.4, angle=270]{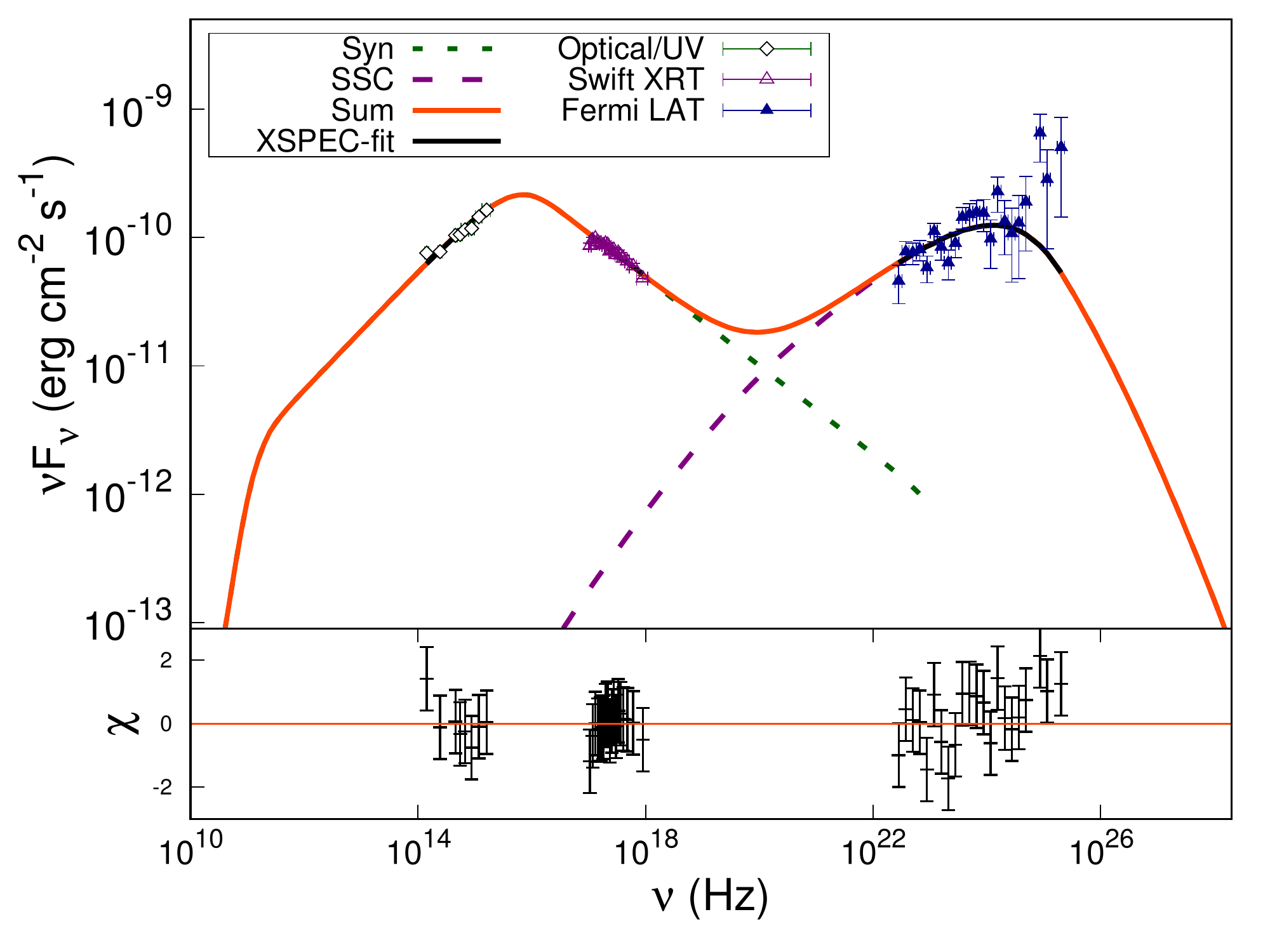}
     }
\vspace*{0.5cm}\caption{One zone leptonic model fits to the broad band SED for epochs A (left figure) and B (right figure) for the source PKS 2155$-$304. The different components in the figure have the same meaning as that of Figure \ref{figure-9}.}
\label{figure-11}
\end{figure*}

To further characterize the physical properties of the sources
during the epochs of optical and $\gamma$-ray flux
variations, we generated the broad band SED of the sources at the different epochs and 
modelled them using the one zone leptonic emission model. 
For comparison, we also generated the broad band SED for a quiescent
state in each of the sources. To generate the SEDs, all
photometric measurements during each epoch as summarized in Table \ref{table-2}
were averaged filter wise in the optical and IR bands to get one photometric 
point for each epoch. For X-rays and $\gamma$-rays, average X-ray and $\gamma$-ray 
spectra were generated using all data during the period of each epoch. Blazars are known to show flux variations over a range of time scales
\citep{1995ARA&A..33..163W}. Also
during the time ranges considered here for SED analysis, the brightness states of the 
sources were not stable in most of the wavelengths. Therefore, the source parameters
obtained by fitting the time averaged SED could be treated as average/typical values
applicable for the duration that is considered.
In the one zone
leptonic emission model the low energy hump of the broad band SED of BL Lacs is due
to synchrotron emission from relativistic electrons in the jet, while
the high energy hump is mostly attributed to inverse Compton emission processes. 
For example in the source
PKS 2155$-$304, \cite{2019arXiv191201880W} explains the high energy part of the 
SED using SSC process. The flare of the source in June 2013 
was well fit by leptonic model, while the flare of April 2013 was fit with
lepto-hadronic model \citep{2020A&A...639A..42A}. Also, the interest in 
the modelling of blazar SED has increased due to the finding of an association
of the IceCube neutrino with blazars such as TXS 0506+056 \citep{2018Sci...361..147I} and
BZB J0955+3551 \citep{2020arXiv200306012P}. In spite of the different model fits attempted
on BL Lac sources such as PKS 2155$-$304  at different periods, we performed a statistical fitting of the broad band SEDs using
synchrotron, synchrotron self Compton and external Compton mechanisms. The details of
the model as implemented within XSPEC can be found in \cite{2018RAA....18...35S}. This
XSPEC implementation of the model also gives the errors in the best fitting parameters
through the $\chi^2$ minimization technique. To account for the model  
as well as observational (for example, uncertainties in host galaxy 
contribution and the optical brightness of the sources) uncertainties, we
added 15\% systematics to the data for all the epochs.
%except the epoch A of  AO 0235+164 which required a systematic of 25\% in order to obtain $\chi^2/dof<2$}. 
The model has twelve free parameters, namely
particle spectral index before the break ($p$), the particle spectral index after
the break ($q$), electron energy density ($U_e$), minimum Lorentz factor of the electrons ($\gamma_{min}$), 
the maximum Lorentz factor of the electrons ($\gamma_{max}$), the break Lorentz
factor of the electron distribution ($\gamma_b$), magnetic field (B), size of the emission 
region (R), bulk Lorentz factor of the jet ($\Gamma$), viewing angle of the jet
($\theta$), the temperature of the external photon field (T), and the fraction of the
external photons that take part in the EC process (f). 
The number of parameters 
defining the model SED is larger than the spectral information extracted from 
observed SED and this forced us to freeze some parameters to typical values and 
perform the fitting procedure. For the target photon filed for EC scattering, we considered the photons from  BLR and torus. The emission from these regions
are assumed to be a blackbody type with the temperature of BLR photons as 42000 K (corresponding to the dominant Ly$\alpha$ line with frequency $2.47\times10^{15}$), while for the photons from the torus we considered a temperature of 800 K. In addition, we enforced equipartition condition between the electron
energy density and magnetic field ($U_e \approx B^2/8/\pi$) which ensures minimum source 
energy \citep{1959ApJ...129..849B}. This imposed 
further constrain on the free parameters. Finally, the fitting was performed on five 
parameters namely $p$, $q$, B, $\Gamma$ and $\gamma_b$; while the remaining seven 
parameters were frozen to values obtained by ``fit by eye" of the quiescent states for all the sources (Table \ref{table-4}). %{\bf Particularly we constrained the $U_e$ such that equipartition is obtained between the emitting particle and magnetic energy density.}
The validity of the fitted parameters will heavily depend on the choice of these 
frozen parameters.  The observed SED along with the model fits are
given in Fig. \ref{figure-9} to Fig. \ref{figure-11}. The best fit model 
parameters are given in Table \ref{table-5}. 
A quick look into the best fit power-law indices of the particle spectrum disfavors the radiative cooling origin of the broken power-law electron distribution. Under this interpretation, one may expect the difference in the power-law index to be $\sim$1 (p and p+1). The corresponding difference in the synchrotron spectral 
index will be $\sim$0.5 \citep{1986rpa..book.....R}. However, the spectral index difference of blazars exceeds this value (Fig 27. of \citealt{2010ApJ...716...30A}). The large difference in particle indices is also seen in case of CGRaBS catalog (Fig. 9 of \citealt{2017ApJ...851...33P}). An alternate explanation for the broken power-law distribution with large index difference could be the presence of multiple acceleration scenarios 
\citep{2008MNRAS.388L..49S}. In addition, the excessively large index difference can also be an artifact introduced by a steeply decaying spectrum. 
For the sources AO 0235+164 and 
OJ 287, during all the epochs, the high energy component is well fit by the combination of
SSC and EC emission proceses. The seed photons for the EC scattering can be from the dusty 
torus and/or the Lyman-$\alpha$ line emission from the BLR. On the other hand, for the source 
PKS 2155$-$304, the high energy component is well fit by the SSC model alone.

\begin{table*}
\hspace{-5cm}
\caption{Values of the parameters that were frozen during the model fits to reproduce the
observed SED. 
%The viewing angle of 2$^{\circ}$ was assumed in all the SED model fits. 
The size of the emission region is in units of 10$^{16}$ cm, the 
temperature of the external photon field T is in Kelvin and the
value of $\gamma_{max}$ is in units of 10$^6$.}
\label{table-4}
%\addtolength{\tabcolsep}{-3.4pt}
\begin{adjustbox}{width=0.7\textwidth,center=\textwidth}
\begin{tabular}{lcccccr} \hline
Object          & R                & $\gamma_{min}$ & $\gamma_{max}$          & $\theta$  & $T_{IR}/T_{BLR}$     & $f_{IR}/f_{BLR}$          \\ 
                &                  &                &                         &             & (K)             &             \\ \hline
AO 0235+164     & $1.32\times10^{19}$           &   50          & 10  & 2        & 800/42000       & $5.76\times10^{-6}$/$1.26\times10^{-10}$    \\ 
OJ 287          &  $1.11\times10^{16}$           &       100      & 10  &  2.0        &    800/--       &   0.05/--     \\
PKS 2155$-$304  & $2.24 \times10^{16}$            &   50           & 100 &  2.0       &  --/--        &   --/--      \\
\hline
\end{tabular}
\end{adjustbox}
\end{table*}

\begin{table*}
\caption{Results of the broad band SED analysis for the sources at different epochs. The subscript and superscript on the parameters are their lower and upper bounds calculated at the 90\% confidence level. A $--$ implies that the upper or lower bound value on the parameter is not constrained.}
\label{table-5}
\begin{tabular}{cccccccc} \hline
      &       & Bulk Lorentz & Low energy     & High energy    & Break energy     & Magnetic     &              \\
Name  & Epoch &  factor      & particle index & particle index & $\gamma_b$ & field (Gauss) &  $\chi^2$/dof \\
\hline
AO 0235+164 &  A  & $82.02_{79.40}^{84.66}$    & $1.20_{--}^{1.31}$  & $7.17_{6.67}^{7.69}$  & $11671_{10373}^{12793}$ & $0.0108_{0.01}^{0.011}$  & 28.71/34 \\ 
            &  B  & $86.26_{83.82}^{88.70}$  & $1.23^{2.75}_{--}$ & $5.93^{6.23}_{5.29}$  & $9055^{10184}_{8132}$ & $0.0112_{0.0106}^{0.0119}$ & 44.26/35 \\
            &  C  & $53.93^{63.85}_{--}$   &  $2.88^{3.07}_{--}$  & $8.00_{6.32}^{8.00}$  & $20082_{16801}^{22052}$ & $0.0082^{--}_{0.0070}$  &  20.84/15 \\
            &  D  & $67.63^{--}_{--}$   &  $2.43^{2.74}_{1.12}$  & $5.70_{5.29}^{6.23}$  & $16410^{--}_{11426}$ & $0.0089^{--}_{0.0075}$  &  12.74/16\\
\hline            
OJ 287      &  A  &  $14.37_{11.50}^{--}$ & $2.06_{1.83}^{2.25}$   & $5.40_{4.89}^{6.37}$  & $2480_{2140}^{--}$  & $0.87_{0.81}^{--}$  &  17.38/16  \\
            &  B  &  $39.62_{15.66}^{--}$    & $1.90_{1.67}^{2.10}$  & $5.18_{4.86}^{5.86}$  & $2168_{1905}^{2504}$  & $0.89_{0.84}^{--}$ &  20.04/25 \\
            &  C  &   $17.44_{14.45}^{21.76}$    & $2.89_{2.76}^{3.02}$  & $7.25_{4.97}^{--}$  & $3963_{2875}^{4852}$  & $1.01_{0.96}^{1.08}$  &  19.90/24 \\
            &  D  &   $28.64_{18.86}^{46.14}$    & $2.29_{2.13}^{2.35}$  & $6.39_{5.71}^{--}$  & $4429_{4141}^{6373}$  & $0.79_{0.77}^{0.86}$  &  20.67/28 \\
\hline
PKS 2155-304 & A  & $10.57_{9.66}^{11.98}$   & $2.33_{2.31}^{2.38}$  & $12_{8.91}^{--}$  & $21658_{17618}^{22767}$ & $0.45_{0.42}^{0.49}$ &  32.86/29 \\
             & B  & $9.15_{7.77}^{--}$   & $2.07_{--}^{2.30}$  & $3.69_{3.50}^{3.92}$  & $15615_{11408}^{20993}$ & $0.49_{0.39}^{--}$ &  26.43/43 \\
\hline
\end{tabular}
\end{table*}

\section{Discussion}
%\subsection{Connection between optical and GeV flux variations}
The nature of seed photons that participate in the IC process to explain the high energy
emission in BL Lac objects is highly debated. Cross-correlation analysis between variations in the
optical band and Gev $\gamma$-ray band on a large sample of blazars tend to support the scenario of
EC to be the dominant process in FSRQs and SSC being the dominant process in 
BL Lacs \citep{2014ApJ...797..137C}. On analysis of the relation between optical 
and $\gamma$-ray flux variations on a large sample of blazars, \cite{2014MNRAS.439..690H} 
found SSC to be the dominant high energy emission mechanism
in ISP and HSP sources, while EC is more dominant in LSP sources. Though these studies broadly support
the one-zone leptonic emission from blazar jets, the recent observations of uncorrelated optical
and GeV flares challenge one zone models of blazar emission. The existence of such uncorrelated flux
variations between the optical and GeV bands are known today owing to the capabilities of {\it Fermi} and
supporting ground based observations at longer wavelengths in the optical bands. Earlier reports on such
uncorrelated optical and GeV flares available in the literature are mostly for the FSRQ type of blazars such
as PKS 0208$-$512 \citep{2013ApJ...763L..11C}, PKS 2142$-$75 \citep{2013ApJ...779..174D}, PKS 0454$-$234, S4 1849+67, BZQ J0850$-$1213, OP 313
\citep{2014ApJ...797..137C}, PKS 1510$-$089 \citep{2015ApJ...804..111M},
3C 454.3 \citep{2019MNRAS.486.1781R}, 3C 273, 3C 279 and CTA 102 
\citep{2020MNRAS.tmp.2558R}. The only BL Lac
object for which uncorrelated optical and GeV flux variations are known is 
PKS 2155$-$304 \citep{2019arXiv191201880W}. We have examined
here on the prevalence of uncorrelated optical and GeV flux variations in the BL Lac objects. Based
on the criteria outlined in Section 2, we arrived at a sample of three objects namely
A0 0235+164, OJ 287 and PKS 2155$-$304. In all the three objects 
in the epochs where flares are studied in this work, the optical and 
$\gamma$-ray flux variations are correlated. However, some observations
point to scenarios of uncorrelated optical and GeV flux variations
to be a common feature in blazars. Recently, from a discrete correlation
function analysis of a sample of 178 blazars, \cite{2019ApJ...880...32L} 
found that about 50\% of optical flares have no $\gamma$-ray counterparts and
about 20\% of $\gamma$-ray flares have no optical counterparts. This has 
increased our knowledge on the prevalence of correlated optical and 
$\gamma$-ray flux variations in blazars manifold compared to the earlier reports 
available on few individual sources 
\citep{2013ApJ...763L..11C,2013ApJ...779..174D,2014ApJ...797..137C,2015ApJ...804..111M,2019MNRAS.486.1781R,2020MNRAS.tmp.2558R}. 

In AO 0235+164, during epochs A, B and D we found variations in the optical and GeV $\gamma$-rays are closely correlated. This is a LSP source \citep{2015ApJ...810...14A} and the high energy emission in the broad band
SEDs during all the epochs of this source is well fit by EC process. 
For epochs C and D the $\gamma$-ray emission can be well fit by considering EC scattering of IR photons from the 
	dusty torous; however, in case of epochs A and B additional EC component scattering off the Lyman-$\alpha$ photons from the BLR is required to obtain a reasonable reduced $\chi^2$( see Fig. \ref{figure-9}).
%{\bf For epochs A and B a two component EC process with seed photons from IR torus and BLR region is required to obtain reasonable reduced $\chi^2$;  while for epocs C and D, a single component EC process with seed photons from IR torus provides a good fit( see Fig. \ref{figure-9}). }
%The seed photons for inverse Compton scattering being the IR emission from the torus (. 
The high energy emission is via EC process is also studied by \cite{2014MNRAS.439..690H}. The multi-wavelength analysis of the source AO 0235+164 conducted by \cite{2017MNRAS.464.4875B} also suggests the presence of EC process, which is responsible for the high energy emission. The
radiation output from inverse Compton emission, that
constitutes the high energy component is a function of the energy density of the electrons $U_e$,
Lorentz factor $\Gamma$ and the density of the external photon field that
participates in IC scattering. At the same time, the radiation output from synchrotron process
that constitutes the low energy component in the SED depends on $U_e$, $\Gamma$ and magnetic field
B. Broadband SED model fits during  different flux states of AO\,0235+164 showed an increasing trend in $\Gamma$ and B, while decreasing trend was noticed in $\gamma_b$ from quiescent to high flux states (see Table \ref{table-5}). However, the large errors or unconstrained upper/lower bounds on the parameters did not lead us to unambiguously conclude that their variations could lead to correlated emission between the optical and $\gamma$-rays.

% The bulk Lorentz factor is relatively larger during  epochs A and B compared to the quiescent period. At the same time the break energy energy is shifted to lower energy during epochs A and B  relative to the quiescent state.  This increase in the bulk Lorentz factor and the shift in the peak energy may manifest as  increase in optical and $\gamma$-ray emission. Thus the correlated optical and $\gamma$-ray flux variations during epochs A and B can be attributed to the increase of $\Gamma$ and shift in $\gamma_b$. At epoch D, we have an optical flare with a $\gamma$-ray counterpart. Our leptonic model fit showed an increase in $\Gamma$ and decrease in $\gamma_b$ during this epoch, relative to quiescent period. Additionally, the magnetic field increase marginally in epoch D as compared to the quiescent state. Hence, the correlated optical and $\gamma$-ray flare seen during epoch D can be inferred as a result of increase in $\Gamma$ and B,  and decrease in $\gamma_b$.

OJ 287 is a LSP BL Lac and here too the high energy part of the SED at different
epochs is described  by EC process with the seed photons from the dusty torus.
In this source, we found during all the flaring epochs
considered here the optical flare is
correlated with the $\gamma$-ray flare (see Figs. \ref{figure-3} and 
\ref{figure-4}). The best fit broadband SED  model parameters obtained in the different flux states of OJ 287 are shown in Table \ref{table-5}. These parameters show increasing/decreasing trend from low to high flux states, however,  due to large error or unconstrained upper/lower bounds on the parameters, we cannot confirm the exact cause of variability in the optical and $\gamma$-ray emission.

For the source PKS 2155$-$304, we identified two epochs, epoch A, a quiescent period and epoch B, an active
period  with an increased optical state coinciding with a $\gamma$-ray flare
(see Fig. \ref{figure-5}). 
This is a HSP BL Lac and the high energy emission in the broad
band SED both during the quiescent and flaring state is fit by SSC process in our one-zone leptonic
modelling approach. 
This source has also been
extensively studied by various authors for multi-wavelength variability and broad band SED modelling,
however, different processes have been invoked at different periods of the source to explain the
observations. For example multi-wavelength observations carried out on the source during 25 August
2008 - 06 September 2008, \cite{2009ApJ...696L.150A} found correlation between the optical brightness
changes with the changes in the VHE $\gamma$-rays, but the optical variations did not correlate
with the GeV $\gamma$-rays. The authors argue that the population of electrons that were
responsible for optical emission may be different from those responsible for GeV and VHE $\gamma$-rays. From
an analysis of the optical and GeV $\gamma$-rays during the period 2007 to 2009, \citep{2014A&A...571A..39H}
found varied correlations between the optical and GeV $\gamma$-rays. They found instances
of (a) correlation between optical and GeV $\gamma$-rays, (b) anti correlation between optical and
GeV $\gamma$-rays and (c) no-correlation between optical and GeV $\gamma$-rays. In this work too,
during epoch B, we have an enhanced $\gamma$-ray with a peak value of about 13 times the mean brightness at the quiescent level.
Similarly, in the optical too there is an enhancement of about a factor of 2 relative to the quiescent optical
brightness, but available observations lack optical measurements during the peak of the $\gamma$-ray flare.
X-ray flux enhancement too is not coincident with the GeV flare. The parameters obtained from model fits to epoch A and B of PKS 2155$-$304 are  given in Table \ref{table-5}. Due to large as well as unconstrained errors on the parameter, the  physical parameters that could lead to flux enhancement in the source during epoch B could not be ascertained.

\section{Summary}
In an effort to identify correlated as well as uncorrelated flux variations between optical and GeV $\gamma$-ray band in BL Lac objects, we carried out a systematic analysis of flux variations
in three BL Lac objects namely AO 0235+164, OJ 287 and PKS 2155$-$304. We summarize the results of the 
work below
\begin{enumerate}
\item All the three BL Lacs showed correlated variations between optical and 
$\gamma$-rays during the flares analyzed in this work. 
\item The high energy hump of the broad band SED of AO 0235+164 and OJ 287 at all epochs are 
described by inverse Compton scattering of IR photons from the torus and/or the line emission 
from the BLR. For PKS 2155$-$304, the SEDs at
all epochs are fit by synchrotron and synchrotron self Compton process.

\item The instances of correlated flux variations in optical and GeV bands 
as well quiescent epochs in all the three sources are 
explained by the one zone leptonic scenario. %The observed flux variations can be explained by changes in the bulk Lorentz factor, magnetic field and electron density.

\item At all the epochs in the three sources where significant colour variations were observed, we found
a bluer when brighter behaviour.

\end{enumerate}

\section*{Acknowledgments}
We thank the referee for his/her suggestions, which helped the authors to improve the manuscript significantly. Data from the spectropolarimetric monitoring project at the Steward Observatory were used for this work. This paper used the optical/near infrared light curves of SMARTS which are available at www.astro.yale.edu/smarts/glast/home.php. This program is supported by Fermi Guest Investigator grants NNX08AW56G, NNX09AU10G, NNX12AO93G, and NNX15AU81G. This research work has extensively used the High Performance Computing Facility of the Indian Institute of Astrophysics, Bangalore.
% Entry for the table of contents, for this guide only

\section{Data Availability}
The multiwavelength data used in this work are publicly available from the {\it Fermi-LAT{\footnote{https://fermi.gsfc.nasa.gov/ssc/data/access/}}, Swift-XRT and Swift-UVOT{\footnote{https://www.ssdc.asi.it/mmia/index.php?mission=swiftmastr}}, SMARTS{\footnote{http://www.astro.yale.edu/smarts/glast/home.php}} and Steward observatory{\footnote{http://james.as.arizona.edu/$\sim$psmith/Fermi/DATA/individual.html}}}.
\bibliographystyle{mnras}
\bibliography{ref}

\begin{thebibliography}{91}
\expandafter\ifx\csname natexlab\endcsname\relax\def\natexlab#1{#1}\fi

\bibitem[{Abdalla} et~al.(2020){Abdalla}, {Adam}, {Aharonian}
  et~al.]{2020A&A...639A..42A}
{Abdalla} H., {Adam} R., {Aharonian} F., et~al., 2020, \aap, 639, A42

\bibitem[{Abdo} et~al.(2010){Abdo}, {Ackermann}, {Agudo}
  et~al.]{2010ApJ...716...30A}
{Abdo} A.~A., {Ackermann} M., {Agudo} I., et~al., 2010, \apj, 716, 30

\bibitem[{Abdollahi} et~al.(2020){Abdollahi}, {Acero}, {Ackermann}
  et~al.]{2020ApJS..247...33A}
{Abdollahi} S., {Acero} F., {Ackermann} M., et~al., 2020, \apjs, 247, 1, 33

\bibitem[{Acero} et~al.(2015){Acero}, {Ackermann}, {Ajello}
  et~al.]{2015ApJS..218...23A}
{Acero} F., {Ackermann} M., {Ajello} M., et~al., 2015, \apjs, 218, 23

\bibitem[{Ackermann} et~al.(2015){Ackermann}, {Ajello}, {Atwood}
  et~al.]{2015ApJ...810...14A}
{Ackermann} M., {Ajello} M., {Atwood} W.~B., et~al., 2015, \apj, 810, 14

\bibitem[{Ackermann} et~al.(2012){Ackermann}, {Ajello}, {Ballet}
  et~al.]{2012ApJ...751..159A}
{Ackermann} M., {Ajello} M., {Ballet} J., et~al., 2012, \apj, 751, 2, 159

\bibitem[{Agudo} et~al.(2011){Agudo}, {Jorstad}, {Marscher}
  et~al.]{2011ApJ...726L..13A}
{Agudo} I., {Jorstad} S.~G., {Marscher} A.~P., et~al., 2011, \apjl, 726, 1, L13

\bibitem[{Aharonian} et~al.(2009){Aharonian}, {Akhperjanian}, {Anton}
  et~al.]{2009ApJ...696L.150A}
{Aharonian} F., {Akhperjanian} A.~G., {Anton} G., et~al., 2009, \apjl, 696, 2,
  L150

\bibitem[{Aharonian}(2000)]{2000NewA....5..377A}
{Aharonian} F.~A., 2000, \na, 5, 377

\bibitem[{Andruchow} et~al.(2005){Andruchow}, {Romero} \&
  {Cellone}]{2005A&A...442...97A}
{Andruchow} I., {Romero} G.~E., {Cellone} S.~A., 2005, \aap, 442, 1, 97

\bibitem[{Angel} \& {Stockman}(1980)]{1980ARA&A..18..321A}
{Angel} J.~R.~P., {Stockman} H.~S., 1980, \araa, 18, 321

\bibitem[{Antonucci}(1993)]{1993ARA&A..31..473A}
{Antonucci} R., 1993, \araa, 31, 473

\bibitem[{Arnaud}(1996)]{1996ASPC..101...17A}
{Arnaud} K.~A., 1996, in { Astronomical Data Analysis Software and Systems
  V\/}, edited by G.~H. {Jacoby}, J.~{Barnes}, vol. 101 of { Astronomical
  Society of the Pacific Conference Series\/}, ~17

\bibitem[{Baring} et~al.(2017){Baring}, {B{\"o}ttcher} \&
  {Summerlin}]{2017MNRAS.464.4875B}
{Baring} M.~G., {B{\"o}ttcher} M., {Summerlin} E.~J., 2017, \mnras, 464, 4,
  4875

\bibitem[{Bessell}(1979)]{1979PASP...91..589B}
{Bessell} M.~S., 1979, \pasp, 91, 589

\bibitem[{B{\l}a{\.z}ejowski} et~al.(2000){B{\l}a{\.z}ejowski}, {Sikora},
  {Moderski} \& {Madejski}]{2000ApJ...545..107B}
{B{\l}a{\.z}ejowski} M., {Sikora} M., {Moderski} R., {Madejski} G.~M., 2000,
  \apj, 545, 107

\bibitem[{Boettcher} et~al.(1997){Boettcher}, {Mause} \&
  {Schlickeiser}]{1997A&A...324..395B}
{Boettcher} M., {Mause} H., {Schlickeiser} R., 1997, \aap, 324, 395

\bibitem[{Bonning} et~al.(2012){Bonning}, {Urry}, {Bailyn}
  et~al.]{2012ApJ...756...13B}
{Bonning} E., {Urry} C.~M., {Bailyn} C., et~al., 2012, \apj, 756, 13

\bibitem[{Bonning} et~al.(2009){Bonning}, {Bailyn}, {Urry}
  et~al.]{2009ApJ...697L..81B}
{Bonning} E.~W., {Bailyn} C., {Urry} C.~M., et~al., 2009, \apjl, 697, L81

\bibitem[{B{\"o}ttcher}(2007)]{2007Ap&SS.309...95B}
{B{\"o}ttcher} M., 2007, \apss, 309, 95

\bibitem[{B{\"o}ttcher} et~al.(2013){B{\"o}ttcher}, {Reimer}, {Sweeney} \&
  {Prakash}]{2013ApJ...768...54B}
{B{\"o}ttcher} M., {Reimer} A., {Sweeney} K., {Prakash} A., 2013, \apj, 768, 54

\bibitem[{Bowyer} et~al.(1984){Bowyer}, {Brodie}, {Clarke} \&
  {Henry}]{1984ApJ...278L.103B}
{Bowyer} S., {Brodie} J., {Clarke} J.~T., {Henry} J.~P., 1984, \apjl, 278, L103

\bibitem[{Breeveld} et~al.(2011){Breeveld}, {Landsman}, {Holland}, {Roming},
  {Kuin} \& {Page}]{2011AIPC.1358..373B}
{Breeveld} A.~A., {Landsman} W., {Holland} S.~T., {Roming} P., {Kuin} N.~P.~M.,
  {Page} M.~J., 2011, in { American Institute of Physics Conference Series\/},
  edited by J.~E. {McEnery}, J.~L. {Racusin}, N.~{Gehrels}, vol. 1358 of {
  American Institute of Physics Conference Series\/},  373--376

\bibitem[{Burbidge}(1959)]{1959ApJ...129..849B}
{Burbidge} G.~R., 1959, \apj, 129, 849

\bibitem[{Burrows} et~al.(2005){Burrows}, {Hill}, {Nousek}
  et~al.]{2005SSRv..120..165B}
{Burrows} D.~N., {Hill} J.~E., {Nousek} J.~A., et~al., 2005, \ssr, 120, 165

\bibitem[{Chatterjee} et~al.(2013){Chatterjee}, {Fossati}, {Urry}
  et~al.]{2013ApJ...763L..11C}
{Chatterjee} R., {Fossati} G., {Urry} C.~M., et~al., 2013, \apjl, 763, L11

\bibitem[{Cohen} et~al.(2014){Cohen}, {Romani}, {Filippenko}
  et~al.]{2014ApJ...797..137C}
{Cohen} D.~P., {Romani} R.~W., {Filippenko} A.~V., et~al., 2014, \apj, 797, 137

\bibitem[{Cohen} et~al.(1987){Cohen}, {Smith}, {Junkkarinen} \&
  {Burbidge}]{1987ApJ...318..577C}
{Cohen} R.~D., {Smith} H.~E., {Junkkarinen} V.~T., {Burbidge} E.~M., 1987,
  \apj, 318, 577

\bibitem[{Dickel} et~al.(1967){Dickel}, {Yang}, {McVittie} \&
  {Swenson}]{1967AJ.....72..757D}
{Dickel} J.~R., {Yang} K.~S., {McVittie} G.~C., {Swenson} G.~W. J., 1967, \aj,
  72, 757

\bibitem[{Dutka} et~al.(2013){Dutka}, {Ojha}, {Pottschmidt}
  et~al.]{2013ApJ...779..174D}
{Dutka} M.~S., {Ojha} R., {Pottschmidt} K., et~al., 2013, \apj, 779, 174

\bibitem[{Edelson} \& {Krolik}(1988)]{1988ApJ...333..646E}
{Edelson} R.~A., {Krolik} J.~H., 1988, \apj, 333, 646

\bibitem[{Falomo} et~al.(2014){Falomo}, {Pian} \&
  {Treves}]{2014A&ARv..22...73F}
{Falomo} R., {Pian} E., {Treves} A., 2014, \aapr, 22, 73

\bibitem[{Fossati} et~al.(1998){Fossati}, {Maraschi}, {Celotti}, {Comastri} \&
  {Ghisellini}]{1998MNRAS.299..433F}
{Fossati} G., {Maraschi} L., {Celotti} A., {Comastri} A., {Ghisellini} G.,
  1998, \mnras, 299, 433

\bibitem[{Gaur} et~al.(2019){Gaur}, {Gupta}, {Bachev}
  et~al.]{2019MNRAS.484.5633G}
{Gaur} H., {Gupta} A.~C., {Bachev} R., et~al., 2019, \mnras, 484, 4, 5633

\bibitem[{Gehrels} et~al.(2004){Gehrels}, {Chincarini}, {Giommi}
  et~al.]{2004ApJ...611.1005G}
{Gehrels} N., {Chincarini} G., {Giommi} P., et~al., 2004, \apj, 611, 1005

\bibitem[{Ghisellini} \& {Madau}(1996)]{1996MNRAS.280...67G}
{Ghisellini} G., {Madau} P., 1996, \mnras, 280, 67

\bibitem[{Ghisellini} \& {Maraschi}(1989)]{1989ApJ...340..181G}
{Ghisellini} G., {Maraschi} L., 1989, \apj, 340, 181

\bibitem[{Ghisellini} \& {Tavecchio}(2008)]{2008MNRAS.387.1669G}
{Ghisellini} G., {Tavecchio} F., 2008, \mnras, 387, 1669

\bibitem[{Ghisellini} et~al.(2011){Ghisellini}, {Tavecchio}, {Foschini} \&
  {Ghirland a}]{2011MNRAS.414.2674G}
{Ghisellini} G., {Tavecchio} F., {Foschini} L., {Ghirland a} G., 2011, \mnras,
  414, 3, 2674

\bibitem[{H.~E.~S.~S. Collaboration} et~al.(2014){H.~E.~S.~S. Collaboration},
  {Abramowski}, {Aharonian} et~al.]{2014A&A...571A..39H}
{H.~E.~S.~S. Collaboration}, {Abramowski} A., {Aharonian} F., et~al., 2014,
  \aap, 571, A39

\bibitem[{Hagen-Thorn} et~al.(2008){Hagen-Thorn}, {Larionov}, {Jorstad}
  et~al.]{2008ApJ...672...40H}
{Hagen-Thorn} V.~A., {Larionov} V.~M., {Jorstad} S.~G., et~al., 2008, \apj,
  672, 1, 40

\bibitem[{Hewitt} \& {Burbidge}(1980)]{1980ApJS...43...57H}
{Hewitt} A., {Burbidge} G., 1980, \apjs, 43, 57

\bibitem[{Hovatta} et~al.(2014){Hovatta}, {Pavlidou}, {King}
  et~al.]{2014MNRAS.439..690H}
{Hovatta} T., {Pavlidou} V., {King} O.~G., et~al., 2014, \mnras, 439, 1, 690

\bibitem[{IceCube Collaboration} et~al.(2018){IceCube Collaboration},
  {Aartsen}, {Ackermann} et~al.]{2018Sci...361..147I}
{IceCube Collaboration}, {Aartsen} M.~G., {Ackermann} M., et~al., 2018,
  Science, 361, 6398, 147

\bibitem[{Kalberla} et~al.(2005){Kalberla}, {Burton}, {Hartmann}
  et~al.]{2005A&A...440..775K}
{Kalberla} P.~M.~W., {Burton} W.~B., {Hartmann} D., et~al., 2005, \aap, 440,
  775

\bibitem[{Kinman} et~al.(1966){Kinman}, {Lamla} \&
  {Wirtanen}]{1966ApJ...146..964K}
{Kinman} T.~D., {Lamla} E., {Wirtanen} C.~A., 1966, \apj, 146, 964

\bibitem[{Konigl}(1981)]{1981ApJ...243..700K}
{Konigl} A., 1981, \apj, 243, 700

\bibitem[{Kushwaha} et~al.(2018){Kushwaha}, {Gupta}, {Wiita}
  et~al.]{2018MNRAS.473.1145K}
{Kushwaha} P., {Gupta} A.~C., {Wiita} P.~J., et~al., 2018, \mnras, 473, 1, 1145

\bibitem[{Kushwaha} et~al.(2013){Kushwaha}, {Sahayanathan} \&
  {Singh}]{2013MNRAS.433.2380K}
{Kushwaha} P., {Sahayanathan} S., {Singh} K.~P., 2013, \mnras, 433, 3, 2380

\bibitem[{Liodakis} et~al.(2019){Liodakis}, {Romani}, {Filippenko}, {Kocevski}
  \& {Zheng}]{2019ApJ...880...32L}
{Liodakis} I., {Romani} R.~W., {Filippenko} A.~V., {Kocevski} D., {Zheng} W.,
  2019, \apj, 880, 1, 32

\bibitem[{Lynden-Bell}(1969)]{1969Natur.223..690L}
{Lynden-Bell} D., 1969, \nat, 223, 690

\bibitem[{MacDonald} et~al.(2015){MacDonald}, {Marscher}, {Jorstad} \&
  {Joshi}]{2015ApJ...804..111M}
{MacDonald} N.~R., {Marscher} A.~P., {Jorstad} S.~G., {Joshi} M., 2015, \apj,
  804, 111

\bibitem[{Madejski} \& {Sikora}(2016)]{2016ARA&A..54..725M}
{Madejski} G.~G., {Sikora} M., 2016, \araa, 54, 725

\bibitem[{Mannheim}(1993)]{1993A&A...269...67M}
{Mannheim} K., 1993, \aap, 269, 67

\bibitem[{Mao} et~al.(2016){Mao}, {Urry}, {Massaro}, {Paggi}, {Cauteruccio} \&
  {K{\"u}nzel}]{2016ApJS..224...26M}
{Mao} P., {Urry} C.~M., {Massaro} F., {Paggi} A., {Cauteruccio} J.,
  {K{\"u}nzel} S.~R., 2016, \apjs, 224, 26

\bibitem[{Marscher} \& {Gear}(1985)]{1985ApJ...298..114M}
{Marscher} A.~P., {Gear} W.~K., 1985, \apj, 298, 114

\bibitem[{Mastichiadis} \& {Kirk}(1995)]{1995A&A...295..613M}
{Mastichiadis} A., {Kirk} J.~G., 1995, \aap, 295, 613

\bibitem[{Mattox} et~al.(1996){Mattox}, {Bertsch}, {Chiang}
  et~al.]{1996ApJ...461..396M}
{Mattox} J.~R., {Bertsch} D.~L., {Chiang} J., et~al., 1996, \apj, 461, 396

\bibitem[{M{\"u}cke} \& {Protheroe}(2001)]{2001APh....15..121M}
{M{\"u}cke} A., {Protheroe} R.~J., 2001, Astroparticle Physics, 15, 121

\bibitem[{Nolan} et~al.(2012){Nolan}, {Abdo}, {Ackermann}
  et~al.]{2012ApJS..199...31N}
{Nolan} P.~L., {Abdo} A.~A., {Ackermann} M., et~al., 2012, \apjs, 199, 2, 31

\bibitem[{Padovani} \& {Giommi}(1995)]{1995ApJ...444..567P}
{Padovani} P., {Giommi} P., 1995, \apj, 444, 567

\bibitem[{Paliya} et~al.(2020){Paliya}, {B{\"o}ttcher}, {Mar{\'\i}a Del Olmo
  Garc{\'\i}a} et~al.]{2020arXiv200306012P}
{Paliya} V.~S., {B{\"o}ttcher} M., {Mar{\'\i}a Del Olmo Garc{\'\i}a} A.,
  et~al., 2020, arXiv e-prints,  arXiv:2003.06012

\bibitem[{Paliya} et~al.(2016){Paliya}, {Diltz}, {B{\"o}ttcher}, {Stalin} \&
  {Buckley}]{2016ApJ...817...61P}
{Paliya} V.~S., {Diltz} C., {B{\"o}ttcher} M., {Stalin} C.~S., {Buckley} D.,
  2016, \apj, 817, 61

\bibitem[{Paliya} et~al.(2017{\natexlab{a}}){Paliya}, {Marcotulli}, {Ajello}
  et~al.]{2017ApJ...851...33P}
{Paliya} V.~S., {Marcotulli} L., {Ajello} M., et~al., 2017{\natexlab{a}}, \apj,
  851, 1, 33

\bibitem[{Paliya} et~al.(2015){Paliya}, {Sahayanathan} \&
  {Stalin}]{2015ApJ...803...15P}
{Paliya} V.~S., {Sahayanathan} S., {Stalin} C.~S., 2015, \apj, 803, 15

\bibitem[{Paliya} et~al.(2017{\natexlab{b}}){Paliya}, {Stalin}, {Ajello} \&
  {Kaur}]{2017ApJ...844...32P}
{Paliya} V.~S., {Stalin} C.~S., {Ajello} M., {Kaur} A., 2017{\natexlab{b}},
  \apj, 844, 1, 32

\bibitem[{Papadakis} et~al.(2007){Papadakis}, {Villata} \&
  {Raiteri}]{2007A&A...470..857P}
{Papadakis} I.~E., {Villata} M., {Raiteri} C.~M., 2007, \aap, 470, 3, 857

\bibitem[{Raiteri} et~al.(2001){Raiteri}, {Villata}, {Aller}
  et~al.]{2001A&A...377..396R}
{Raiteri} C.~M., {Villata} M., {Aller} H.~D., et~al., 2001, \aap, 377, 396

\bibitem[{Raiteri} et~al.(2009){Raiteri}, {Villata}, {Capetti}
  et~al.]{2009A&A...507..769R}
{Raiteri} C.~M., {Villata} M., {Capetti} A., et~al., 2009, \aap, 507, 769

\bibitem[{Rajput} et~al.(2020){Rajput}, {Stalin} \&
  {Sahayanathan}]{2020MNRAS.tmp.2558R}
{Rajput} B., {Stalin} C.~S., {Sahayanathan} S., 2020, \mnras

\bibitem[{Rajput} et~al.(2019){Rajput}, {Stalin}, {Sahayanathan}, {Rakshit} \&
  {Mandal}]{2019MNRAS.486.1781R}
{Rajput} B., {Stalin} C.~S., {Sahayanathan} S., {Rakshit} S., {Mandal} A.~K.,
  2019, \mnras, 486, 2, 1781

\bibitem[{Rakshit} et~al.(2017){Rakshit}, {Stalin}, {Muneer}, {Neha} \&
  {Paliya}]{2017ApJ...835..275R}
{Rakshit} S., {Stalin} C.~S., {Muneer} S., {Neha} S., {Paliya} V.~S., 2017,
  \apj, 835, 2, 275

\bibitem[{Rybicki} \& {Lightman}(1986)]{1986rpa..book.....R}
{Rybicki} G.~B., {Lightman} A.~P., 1986, {Radiative Processes in Astrophysics}

\bibitem[{Safna} et~al.(2020){Safna}, {Stalin}, {Rakshit} \&
  {Mathew}]{2020MNRAS.tmp.2538S}
{Safna} P.~Z., {Stalin} C.~S., {Rakshit} S., {Mathew} B., 2020, \mnras

\bibitem[{Sahayanathan}(2008)]{2008MNRAS.388L..49S}
{Sahayanathan} S., 2008, \mnras, 388, 1, L49

\bibitem[{Sahayanathan} et~al.(2018){Sahayanathan}, {Sinha} \&
  {Misra}]{2018RAA....18...35S}
{Sahayanathan} S., {Sinha} A., {Misra} R., 2018, Research in Astronomy and
  Astrophysics, 18, 035

\bibitem[{Sarkar} et~al.(2019){Sarkar}, {Chitnis}, {Gupta}
  et~al.]{2019ApJ...887..185S}
{Sarkar} A., {Chitnis} V.~R., {Gupta} A.~C., et~al., 2019, \apj, 887, 2, 185

\bibitem[{Shakura} \& {Sunyaev}(1973)]{1973A&A....24..337S}
{Shakura} N.~I., {Sunyaev} R.~A., 1973, \aap, 24, 337

\bibitem[{Shimmins} \& {Bolton}(1974)]{1974AuJPA..32....1S}
{Shimmins} A.~J., {Bolton} J.~G., 1974, Australian Journal of Physics
  Astrophysical Supplement, 32, 1

\bibitem[{Sillanpaa} et~al.(1988){Sillanpaa}, {Haarala}, {Valtonen},
  {Sundelius} \& {Byrd}]{1988ApJ...325..628S}
{Sillanpaa} A., {Haarala} S., {Valtonen} M.~J., {Sundelius} B., {Byrd} G.~G.,
  1988, \apj, 325, 628

\bibitem[{Smith} et~al.(2009){Smith}, {Montiel}, {Rightley}, {Turner},
  {Schmidt} \& {Jannuzi}]{2009arXiv0912.3621S}
{Smith} P.~S., {Montiel} E., {Rightley} S., {Turner} J., {Schmidt} G.~D.,
  {Jannuzi} B.~T., 2009, ArXiv e-prints

\bibitem[{Spinrad} \& {Smith}(1975)]{1975ApJ...201..275S}
{Spinrad} H., {Smith} H.~E., 1975, \apj, 201, 275

\bibitem[{Stalin} et~al.(2009){Stalin}, {Kawabata}, {Uemura}
  et~al.]{2009MNRAS.399.1357S}
{Stalin} C.~S., {Kawabata} K.~S., {Uemura} M., et~al., 2009, \mnras, 399, 3,
  1357

\bibitem[{Ulrich} et~al.(1997){Ulrich}, {Maraschi} \&
  {Urry}]{1997ARA&A..35..445U}
{Ulrich} M.-H., {Maraschi} L., {Urry} C.~M., 1997, \araa, 35, 445

\bibitem[{Urry} \& {Padovani}(1995)]{1995PASP..107..803U}
{Urry} C.~M., {Padovani} P., 1995, \pasp, 107, 803

\bibitem[{Villata} et~al.(2004){Villata}, {Raiteri}, {Kurtanidze}
  et~al.]{2004A&A...421..103V}
{Villata} M., {Raiteri} C.~M., {Kurtanidze} O.~M., et~al., 2004, \aap, 421, 103

\bibitem[{Wagner} \& {Witzel}(1995)]{1995ARA&A..33..163W}
{Wagner} S.~J., {Witzel} A., 1995, \araa, 33, 163

\bibitem[{Wierzcholska} et~al.(2019){Wierzcholska}, {Zacharias}, {Jankowsky},
  {Wagner} \& {H.~E.~S.~S. Collaboration}]{2019arXiv191201880W}
{Wierzcholska} A., {Zacharias} M., {Jankowsky} F., {Wagner} S., {H.~E.~S.~S.
  Collaboration}, 2019, arXiv e-prints,  arXiv:1912.01880

\bibitem[{Wood} et~al.(2017){Wood}, {Caputo}, {Charles}, {Di Mauro}, {Magill}
  \& {Jeremy Perkins for the Fermi-LAT Collaboration}]{2017arXiv170709551W}
{Wood} M., {Caputo} R., {Charles} E., {Di Mauro} M., {Magill} J., {Jeremy
  Perkins for the Fermi-LAT Collaboration}, 2017, ArXiv e-prints

\bibitem[{Zhang} et~al.(2014){Zhang}, {Zhao}, {Wang} \&
  {Dai}]{2014RAA....14..933Z}
{Zhang} B.-K., {Zhao} X.-Y., {Wang} C.-X., {Dai} B.-Z., 2014, Research in
  Astronomy and Astrophysics, 14, 8, 933

\bibitem[{Zhang} et~al.(2017){Zhang}, {Yan}, {Liao} \&
  {Wang}]{2017ApJ...835..260Z}
{Zhang} P.-f., {Yan} D.-h., {Liao} N.-h., {Wang} J.-c., 2017, \apj, 835, 2, 260

\end{thebibliography}

%%%%%%%%%%%%%%%%%%%%%%%%%%%%%%%%%%%%%%%%%%%%%%%%%%

%%%%%%%%%%%%%%%%% APPENDICES %%%%%%%%%%%%%%%%%%%%%

%%%%%%%%%%%%%%%%%%%%%%%%%%%%%%%%%%%%%%%%%%%%%%%%%%

% Don't change these lines
\bsp	% typesetting comment
\label{lastpage}
\end{document}